\documentclass[fleqn,usenatbib]{mnras}

\usepackage{mathptmx}

\usepackage[T1]{fontenc}
\usepackage{ae,aecompl}

\usepackage{graphicx}	
\usepackage{amsmath}	
\usepackage{amssymb}	

\usepackage{color} 
\usepackage{lineno} 

\providecommand{\myfloor}[1]{\left \lfloor #1 \right \rfloor }
\newcommand{\Ng}{{N_{\rm g}}}
\newcommand{\vk}{{\vec{k}}}
\newcommand{\vq}{{\vec{q}}}
\newcommand{\vr}{{\vec{r}}}
\renewcommand{\d}{{\mathrm{d}}}

\newcommand{\non}{\nonumber}

\let\Re\relax
\DeclareMathOperator{\Re}{Re}
\definecolor{violet}{RGB}{128,0,192}

\definecolor{orange}{RGB}{255,69,0}

\title[Rotation method for sBF integrals]{Rotation method for accelerating multiple-spherical Bessel function integrals against a numerical source function}

\author[Slepian et al.]{Zachary Slepian,$^{1,2,3}$\thanks{E-mail: zslepian@ufl.edu (ZS)}
Yin Li,$^{3,4,7}$\thanks{E-mail: yinli@flatironinstitute.org (YL)}
Marcel Schmittfull,$^{5}$
\& Zvonimir Vlah$^{6}$
\\
$^{1}$Department of Astronomy, University of Florida, 211 Bryant Space Science Center, Gainesville, FL 32611-2055, USA\\
$^{2}$Lawrence Berkeley National Laboratory, 1 Cyclotron Road, Berkeley, CA 94720, USA\\
$^{3}$Berkeley Center for Cosmological Physics, University of California, Berkeley, Berkeley, CA 94720, USA\\
$^{4}$Kavli Institute for Physic and Mathematics of the Universe, 5-1-5 Kashiwanoha, Kashiwa, Chiba Prefecture 277-8583, Japan\\
$^{5}$Institute for Advanced Study, 1 Einstein Drive, Princeton, NJ 08540, USA\\
$^{6}$Centre Europ\'een pour Recherche Nucleaire, Espl. des Particules 1, 1217 Meyrin, Switzerland\\
$^{7}$Center for Computational Astrophysics \& Center for Computational Mathematics, Flatiron Institute, New York, NY, 10010, USA}

\date{Accepted XXX. Received YYY; in original form ZZZ}

\pubyear{2019}

\begin{document}
\label{firstpage}
\pagerange{\pageref{firstpage}--\pageref{lastpage}}
\maketitle

\begin{abstract}
A common problem in cosmology is to integrate the product of two or more spherical Bessel functions (sBFs) with different configuration-space arguments against the power spectrum or its square, weighted by powers of wavenumber.  These integrals generically emerge when correlation functions are evaluated at a displacement from the origin of coordinates. Naively computing them scales as $N_{\rm g}^{p+1}$ with $p$ the number of configuration space arguments and $N_{\rm g}$ the grid size, and they cannot be done with Fast Fourier Transforms (FFTs). Here we show that by rewriting the sBFs as sums of products of sine and cosine and then using the product to sum identities, these integrals can then be performed using 1-D FFTs with $N_{\rm g} \log N_{\rm g}$ scaling. This refactorization is a 45$^{\circ}$ rotation from integrals onto pairs of configuration space arguments (e.g. $(a, b)$) to sums of integrals onto $single$ differences $a-b$ and $a+b$. Hence we call it ``the rotation method.'' It has the potential to accelerate significantly a number of calculations in cosmology, such as perturbation theory predictions of loop integrals, higher order correlation functions, and analytic templates for correlation function covariance matrices. We implement this approach numerically both in a free-standing, publicly-available \textsc{Python} code and within the larger, publicly-available package \texttt{mcfit}. The rotation method evaluated with direct integrations already offers a factor of 6-10$\times$ speed-up over the naive approach in our test cases. Using FFTs, which the rotation method enables, then further improves this to a speed-up of $\sim$$1000-3000\times$ over the naive approach. The rotation method should be useful in light of upcoming large datasets such as DESI or LSST. In analysing these datasets recomputation of these integrals a substantial number of times, for instance to update perturbation theory predictions or covariance matrices as the input linear power spectrum is changed, will be one piece in a Monte Carlo Markov Chain cosmological parameter search: thus the overall savings from our method should be significant. We make our \textsc{python} code publicly available at \url{https://github.com/eelregit/sbf_rotation}
\end{abstract}




\section{Introduction}

Fourier space and configuration space are the two complementary bases
for most problems in cosmology. With the exception of radio, most
observations are done in configuration space; for example, a map of
the 3-D positions of galaxies or of Cosmic Microwave Background (CMB)
temperature as a function of angle on the sky. However, many of the
calculations are simpler in Fourier space. Advantages include that
the Fourier Transform (FT) converts spatial derivatives into multiplication by powers of the wavevector, and that translation invariance is built in since for plane waves
it introduces only a trivial phase factor.\footnote{This is just Bloch's theorem. Furthermore, we note that plane waves are eigenstates of momentum, the Noether's theorem conserved quantity associated with translation invariance.}
Moreover, for a Gaussian Random Field quantities like the power spectrum
are diagonal in the wave-vector magnitude $|\vec{k}|$, and the covariance matrix is often simpler in Fourier space.\footnote{By Gaussian Random Field we mean that the real and imaginary parts of each Fourier-space mode are drawn from a Gaussian at each $k$ with variance given by the power spectrum $P(k)$. This results in the complex phase's being uniformly distributed from $0$ to $2\pi$.} Further, at the level of linear perturbation
theory, all Fourier modes evolve independently (e.g. \citealt{Bernardeau:2002}).

Given these complementary advantages to the two spaces, one often needs
to convert between them, which can be computationally expensive in some cases,
e.g.~for higher-order correlation functions and covariances. An added complication is that, while Fourier modes are well-suited
to a Cartesian grid, problems in cosmology often invoke a degree of
isotropy about the observer, making spherical coordinates a more natural
basis than Cartesian.\footnote{We write ``degree'' of isotropy because when working in redshift-space one does not have full rotational symmetry, though one retains azimuthal symmetry about the line of sight. Furthermore, the RSD are in fact isotropic about the point from which the galaxy survey is observed, just not any other point.} Thus not only does one require an efficient
way to transform between Fourier and configuration spaces, but one
would like to end up in spherical coordinates.

For Fourier transforms from Cartesian wavevectors to Cartesian positions,
Fast Fourier Transform (FFT) algorithms have long been available that
reduce what is naively (for a 1-D transform) an $N_{\rm g}^{2}$ problem to
$\Ng\log \Ng$, where $\Ng$ is the number of grid points in configuration
space or Fourier space \citep{Cooley:1965}. The simplest
versions of these algorithms use the Danielson-Lanczos lemma \citep{Danielson:1942} to bisect the transform into even and odd pieces
of length $N_{\rm g}/2$, meaning the full result is given by a single complex transform
of length $N_{\rm g}/2$. The bisection can be continued until one has expressed
the desired transform as a sum over $c$ transforms of length unity,
where $c$ is the number of bisections. The factor of $\log N_{\rm g}$ thus
comes from the number of bisections $N_{\rm b}$ required, given by $2^{N_{\rm b}}=N_{\rm g}\to N_{\rm b}\sim \log N_{\rm g}$.
This is why the simplest FFT requires that the input be a power of
2. Modern implementations relax this restriction.

For the spherical analog, one can use the plane wave expansion into
spherical Bessel functions and spherical harmonics to convert the
plane waves into spherical coordinates (e.g. \citealt{AWH13} equation 16.61). Often the angular integrals can then be performed analytically; many problems in cosmology have
relatively simple angular structure or indeed have full spherical
symmetry. This integration leaves a 1-D integral from a 3-D wave vector's
magnitude $k=|\vec{k}|$$ $ to radius $r$ against a spherical Bessel
function $j_{\ell}(kr).$ This integral is essentially a Hankel transform
up to details of the integration measure, which can be absorbed into
the function being transformed if desired.

Many fast algorithms exist for the Hankel transform. Most exploit
the change of variables of \cite{Siegman:77} where one sets $kr=\exp\left[\ln k+\ln r\right]$
to render the transform a convolution in $\ln k$ or $\ln r$ that
can then be performed using an FFT. These algorithms are further discussed in \cite{Hamilton:2000}.

However, these algorithms treat only a bijective mapping from $k$
to $r.$ Yet in cosmology an additional case that often arises is
a 1-D integral over $k$ involving a product of spherical Bessel functions
with differing configuration space arguments, e.g. $(2\pi)^{-2}\int k^{2}dk\; P(k)j_{\ell}(kr)j_{\ell'}(kr')$,
where $P(k)$ is the function being transformed, often the power spectrum,
and $j_{\ell}$ is a spherical Bessel function of order $\ell$. It is not
obvious how to evaluate this integral using 1-D FFTs.

In this work, we show how to reduce this computation to a sum over
1-D FFTs. Indeed, we show that the method we present for a two spherical-Bessel-function
integral generalizes to integrals over an arbitrary number of products
of spherical Bessel functions, and we include examples for
the case of three spherical Bessel functions as well.

Our approach should substantially accelerate calculations involving
these integrals, which are often a computational bottleneck. A number of examples exist: exact projection of the 3-D power spectrum to the angular power  spectrum without invoking the Limber approximation \citep{Limber:1953}, computation of loop integrals for the correlation function or power spectrum
in cosmological perturbation theory (PT) (\citealt{Schmittfull_2loop:2016}, \citealt{Slepian_PT:2018}), computation of predictions
for the tree-level isotropically-averaged redshift-space 3-point correlation
function (3PCF) as well as the anisotropic analog; computation of
the 2PCF, 3PCF, and higher correlation functions' covariances in the Gaussian-Random-Field
limit (\citealt{Xu:2013}, \citealt{SE_3pt_alg}, \citealt{SE_aniso_3pt}, Slepian, Cahn \& Eisenstein in prep.).

Developing new, faster tools for these kinds of calculations is particularly vital in the context of the wealth of current and emerging large-scale structure datasets with millions to billions of objects. These datasets will offer unprecedented precision, frequently sub-percent on any given parameter, and thus higher precision predictions and weighting of the data will be critical. In particular, relevant current and upcoming efforts are the Sloan Digital Sky Survey (SDSS; e.g. \citealt{Alam:2017}), WiggleZ \citep{WiggleZ:2012}, and Vipers \citep{Vipers:2018}, as well as upcoming,
such as extended Baryon Oscillation Spectroscopic Survey (eBOSS; \citealt{Dawson:2016}), Dark Energy Spectroscopic Instrument (DESI\footnote{\url{https://www.desi.lbl.gov/}}; \citealt{DESI:2016}) and Euclid\footnote{\url{https://www.euclid-ec.org/}} \citep{Euclid:2011}. There are also significant photometric
datasets, both now available, such as Dark Energy Survey (DES\footnote{\url{https://www.darkenergysurvey.org/}}; \citealt{DES:2005}), and upcoming, such as Large Synoptic Survey Telescope (LSST\footnote{\url{https://www.lsst.org/}}; \citealt{LSST:2012}).
There will also be much intensity mapping data, e.g. from CHIME\footnote{\url{https://chime-experiment.ca/}} \citep{Bandura:2014}, SKA\footnote{\url{https://www.skatelescope.org/}} \citep{SKA_redbook}, and HIRAX\footnote{\url{https://hirax.ukzn.ac.za/}} \citep{Newburgh:2016} along with envisioned future experiments such as PUMA\footnote{\url{https://www.puma.bnl.gov/}} \citep{puma} and others \citep{Hwp}.

To fully exploit these datasets, clustering analyses will need to be faster
than ever, and explore a larger space of variations, such as in the
cosmological parameters. In particular, a common recent approach is
to do Monte Carlo Markov Chain (MCMC) exploration over several cosmological
parameters at a time, requiring a potentially substantial number of
these integrals (see e.g. \citealt{Cataneo:2017} for discussion of how to do this with a Taylor series about a given cosmology). Being able to evaluate theory predictions quickly will
enable this exploration. In particular, fast computation of PT loop
integrals will allow exploiting these datasets down to smaller scales, important for constraining modified gravity models (e.g. DESI collaboration Modified Gravity whitepaper, in prep.) and galaxy biasing models (\citealt{McDonald:2009}, \citealt{Assassi:2014}, \citealt{Senatore:2015}, \citealt{Mirbabayi:2015}, \citealt{Angulo:2015}; for a review, \citealt{Desjacques:2018}).

Further, efficient analytic evaluation of covariance matrices will
be important for optimal weighting of these datasets while avoiding
obtaining the covariance from a large number of mocks, which is computationally expensive (\citealt{Percival:2014}), although alternatives to reduce the cost have been proposed (\citealt{Xu:2013}, \citealt{SE_3pt_alg}, \citealt{Padmanabhan:2016}, \citealt{OConnell:2016}, \citealt{Mohammed:2017}, \citealt{Howlett:2017}, \citealt{Slepian:2018}, \citealt{Li:2018}). Many of these alternatives require many of the integrals we treat in the current work.

This paper is laid out as follows. In \S\ref{sec:emergence}, we present the simplest
case in which double sBF integrals of the power spectrum emerge, from multipole expansion of correlation functions evaluated
with both endpoints offset from the origin. In \S\ref{sec:examples} we present a number of the cosmological use cases where they enter, and focus on presenting the anisotropic 2-Point Correlation Function (2PCF) covariance in more detail in \S\ref{sec:Covar}. In \S\ref{sec:technique} we outline
our technique for evaluating these integrals. \S\ref{sec:general} generalizes the
technique to integrals against three or more spherical Bessel functions. In \S\ref{sec:accel} we discuss
a further acceleration idea generalizing recent work by \cite{Assassi:2017}, \cite{Gebhardt:2018}, and \cite{Simonovic:2018}. We present numerical tests of these ideas in \S\ref{sec:imp}. We conclude in \S\ref{sec:concs}.

\section{How these integrals emerge}
\label{sec:emergence}
Consider the usual transformation between the power spectrum $P(k)$ and
the correlation function $\xi(r)$. Our convention will be to use
a minus sign in the exponential for a 3-D inverse FT and normalize
it as $d^{3}\vec{k}/\left(2\pi\right)^{3}.$ We have
\begin{equation}
\xi(r)=\int \frac{d\Omega_r}{4\pi} \int\frac{d^{3}\vec{k}}{\left(2\pi\right)^{3}}e^{-i\vec{k}\cdot\vec{r}}P(k),
\end{equation}
with $r\equiv|\vec{r}|$. We rewrite the plane wave in terms of spherical Bessel functions $j_{\ell}$ and spherical harmonics $Y_{\ell m}$ using the plane wave expansion (e.g. \citealt{AWH13} equation 16.61),
\begin{equation}
e^{-i\vec{k}\cdot\vec{r}}=4\pi\sum_{\ell m}\left(-i\right)^{\ell}j_{\ell}(kr)Y_{\ell m}(\hat{k})Y_{\ell m}^{*}(\hat{r}).
\label{eqn:pwe}
\end{equation}
We can integrate over $d\Omega_{k}$ since the power spectrum has
no angular dependence. This sets $\ell m = 00$ by orthogonality of the
spherical harmonics. We also perform the integral over $d\Omega_r$. Our integral becomes
\begin{equation}
\xi(r)=\int\frac{k^{2}dk}{2\pi^{2}}j_{0}(kr)P(k).
\label{eqn:basic_xi}
\end{equation}
Naively, computing this integral will scale as $N_r N_k$ if one uses $N_r$ sample points in $r$ and $N_k$ sample points in $k$. This becomes order-$\Ng^2$ if one uses an equal number of grid points, $\Ng$, in each.

However, this integral can be accelerated using a 1-D FFT by setting $kr=\exp\left[\ln k+\ln r\right]$, a transformation first presented in \cite{Siegman:77}. Defining $u=\ln k$ and $v=\ln r$ yields
\begin{equation}
\xi(r)=\int\frac{e^{3u}du}{2\pi^{2}}j_{0}(e^{u+v})P(e^{u})=\bigg\{ \left[e^{3u}P(e^{u})\right]\star\left[j_{0}(e^{u})\right]\bigg\} (v(r)),
\label{eqn:xi_u_v}
\end{equation}
where ``star'' means convolution. The change of variables to $d \ln k$ required one extra power of $e^u$ in the Jacobian: $k^2 dk = (e^u)^2 e^u du$ since $du  = dk/k$. Our notation in the second equality reflects that the convolution is an integral  of the convolvands over the dummy variable $u$ offset from each other by the free variable $v$, which depends on $r$. By the Convolution Theorem equation (\ref{eqn:xi_u_v}) becomes
\begin{align}
&\bigg\{ \left[e^{3u}P(e^{u})\right]\star\left[j_{0}(e^{u})\right]\bigg\} (v)=\nonumber\\
&{\rm FT}^{-1}\bigg\{ {\rm FT}\left[e^{3u}P(e^{u})\right](s){\rm FT}\left[j_{0}(e^{u})\right](s)\bigg\} (v),
\label{eqn:FTs}
\end{align}
where our notation is that the FT is an integral over the dummy variable $u$ evaluated at the free variable $s$, and then the inverse FT integrates over $s$ to be evaluated at $v$. Equation  (\ref{eqn:FTs}) allows use of FFTs to obtain the results at all $v$ in $N_{\rm g}\log N_{\rm g}$
time. We note that this approach appears to require a one-to-one match between
configuration space and Fourier space variables: there is just one
$k$ and one $r$, or equivalently $u$ and $v.$

Now consider a correlation function offset from the origin by some
vector $\vec{r}'$:
\begin{equation}
\xi(|\vec{r}+\vec{r}'|)=\int\frac{d^{3}\vec{k}}{\left(2\pi\right)^{3}}e^{-i\vec{k}\cdot\left(\vec{r}+\vec{r}'\right)}P(k).
\end{equation}
There are two ways to expand this into separated radial and angular
integrals. The first is to treat $\vec{r}+\vec{r}'$ as a unit and
use equation (\ref{eqn:basic_xi}) with $r \to |\vec{r} + \vec{r}'|$:
\begin{equation}
\xi\left(|\vec{r}+\vec{r}'|\right)=\int\frac{k^{2}dk}{2\pi^{2}}j_{0}(k|\vec{r}+\vec{r}'|)P(k).
\label{eqn:twoD_decomp}
\end{equation}
This seems to significantly increase the computational burden because
now one has three relevant parameters: $r,\; r',$ and $\hat{r}\cdot\hat{r}'$, as $|\vec{r} + \vec{r}'| = \sqrt{r^2 + r'^2 + 2r r' \hat{r}\cdot\hat{r}'}$.

The second option is to factor the exponential as
\begin{equation}
e^{-i\vec{k}\cdot\left(\vec{r}+\vec{r}'\right)}=e^{-i\vec{k}\cdot\vec{r}}e^{-i\vec{k}\cdot\vec{r}'}
\end{equation}
and then expand each exponential using the plane wave expansion (\ref{eqn:pwe}) twice. Doing so we obtain
\begin{align}
&\xi\left(|\vec{r}+\vec{r}'|\right)=\left(4\pi\right)^{2}\sum_{\ell \ell'}\left(-i\right)^{\ell+\ell'}\sum_{m m'}Y_{\ell m}(\hat{r})Y_{\ell' m'}^{*}(\hat{r}')\nonumber\\
&\times\int\frac{k^{2}dk}{2\pi^{2}}j_{\ell}(kr)j_{\ell'}(kr')P(k)\int\frac{d\Omega_{k}}{4\pi}Y_{\ell m}^{*}(\hat{k})Y_{\ell' m'}(\hat{k}).
\end{align}
Using orthogonality to perform the angular integral over $d\Omega_k$ we find
\begin{align}
&\xi\left(|\vec{r}+\vec{r}'|\right)=4\pi\sum_{\ell}\left(-1\right)^{\ell}\sum_{m}Y_{\ell m}(\hat{r})Y_{\ell m}^{*}(\hat{r}')\nonumber\\
&\times \int\frac{k^{2}dk}{2\pi^{2}}j_{\ell}(kr)j_{\ell}(kr')P(k)\nonumber\\
&=\sum_{\ell} \left(-1\right)^{\ell}\left(2\ell+1\right)\mathcal{L}_{\ell}(\hat{r}\cdot\hat{r}')\int\frac{k^{2}dk}{2\pi^{2}}j_{\ell}(kr)j_{\ell}(kr')P(k),
\label{eqn:xi_series}
\end{align}
where $\mathcal{L}_{\ell}$ is a Legendre polynomial of degree $\ell$,
and we used the spherical harmonic addition theorem (e.g. \citealt{AWH13} equation 16.57)
\begin{align}
\mathcal{L}_{\ell}(\hat{r} \cdot \hat{r}') = \frac{4\pi}{2\ell+1}\sum_{m=-\ell}^{\ell} Y_{\ell m}(\hat{r})Y_{\ell m}^*(\hat{r}')
\label{eqn:second_reduction}
\end{align}
to resum the spherical harmonics over spins $m$ and obtain the second equality in equation (\ref{eqn:xi_series}).
As expected, the result (\ref{eqn:xi_series}) depends on three parameters: the magnitudes of
the two vectors $r$ and $r'$ and the angle between them $\hat{r}\cdot\hat{r}'\equiv\mu$.
Before (equation \ref{eqn:twoD_decomp}) the integral was over a 3-D grid in $r,r'$ and angle cosine $\mu = \hat{r}\cdot\hat{r}'$; now
we have traded this for an infinite sum of 2-D-grid integrals onto $r$ and $r'$. We note
also that a given number $N_{\mu}$ of grid points in angle cosine would track a maximum frequency $N_{\mu}$ corresponding to a maximal
multipole $\ell_{{\rm max}}=N_{\mu}$ because $\mathcal{L}_{\ell}\propto\cos^{\ell}\theta\propto\cos \ell\theta$---so
the trade is a fair one. Rather than tracking $N_{\mu}$ angle cosines over which to integrate, we must track this number of multipole coefficients in our series for the correlation function.

Moving forward, we define the integral in equation (\ref{eqn:xi_series}) as the $\ell = \ell'$, $n=0$
limit of the more general
\begin{equation}
f_{\ell \ell'}^n(r,r')=\int\frac{k^{2}dk}{2\pi^{2}} k^n j_{\ell}(kr)j_{\ell'}(kr')P(k).
\label{eqn:f_tensor}
\end{equation}
We see from equation (\ref{eqn:xi_series}) that the physical interpretation of these integrals
is as the radial expansion coefficients in a multipole series for the correlation
function shifted from the origin by $\vec{r}'$. If there is an external angular momentum in the problem, such as the quadrupolar modulations that linear-theory redshift space distortions (RSD) induce (e.g. \citealt{Hamilton:1992}), then an $f$-tensor with three sBFs may also arise, defined as
\begin{equation}
f_{\ell \ell' \ell''}^n(r,r',r'')=\int\frac{k^{2}dk}{2\pi^{2}} k^n j_{\ell}(kr)j_{\ell'}(kr')j_{\ell''}(kr'')P(k).
\label{eqn:f_3_tensor}
\end{equation}
The
2-D integral representation involving $j_{0}(k|\vec{r}+\vec{r}'|)$ (equation \ref{eqn:twoD_decomp})
and the infinite sum representation (equation \ref{eqn:xi_series}) are equivalent in principle. However as examples in the following section will show, one often needs only
a finite, small set of these tensors. Consequently in
practice, for such applications the second, factorized representation (\ref{eqn:xi_series})
offers a considerable computational acceleration over the first, unfactored
representation.

We note that multipole coefficients could also be obtained after the fact from the first approach equation (\ref{eqn:twoD_decomp}), but the total work would be larger than in the second approach equation (\ref{eqn:xi_series}). The first would require an $N_{\mu}\times N_{r}\times N_{r'}$
integral and then 1-D projection integrals onto $\mathcal{L}_{\ell}(\hat{r}\cdot\hat{r}')$
for, say, $N_{\ell}$ desired $\ell$, making the total work scale as $N_{\mu}^{2}N_{r}N_{r'}N_{\ell}$. In contrast, in the second approach one needs to do $N_{\ell}$ 2-D integrals
on a grid $N_{r}\times N_{r'}$, for total work scaling as $N_{\ell}N_{r}N_{r'}$. The fact that the work using these two approaches to obtain multipole coefficients differs by a factor of $N_{\mu}$ is sensible: the advantage of the second method is that it performs all the angular integrals analytically. This of course also confers an improvement in the expected precision.

\section{Examples}
\label{sec:examples}
In this section we present a number of cosmological use-cases for the integrals treated in this paper.
\subsection{Exact Angular Power Spectra}
Angular power spectra of galaxy clustering or lensing measurements can be computed by projecting the 3-D matter power spectrum $P(k,z)$ onto the sky, where $z$ is the redshift.
This 3-D to 2-D projection yields a line-of-sight integral over two spherical Bessel functions, two redshift kernels, and the 3-D matter power spectrum (e.g. \citealt{Assassi:2017, Gebhardt:2018}).

Usually, these integrals are simplified using the Limber approximation (\citealt{Limber:1953}), but this breaks down on large scales, corresponding to low multipoles $\ell\lesssim 10-20$.
When modeling angular statistics on such large scales one therefore needs to perform full line-of-sight integrals with two spherical Bessel functions, corresponding to the general type of integral we are evaluating in this paper. Our method at present can already be used for the first few beyond-Limber terms, where the correction is most important. In principle, it can handle the higher-$\ell$ corrections as well. However in practice, due to numerical challenges in canceling divergences in our splitting of the integrals, our current implementation does not extend to the $\ell \sim 10-20$ ultimately desirable for beyond-Limber.

\subsection{Loop Corrections to the Power Spectrum}
In Eulerian Standard Perturbation Theory (SPT) the leading order correction to the linear power spectrum arises from 1-loop integrals. These can be evaluated using a truncated Fourier series representation of the power spectrum \citep{McEwen:2016} or direct representation in terms of 1-D spherical Bessel integrals \citep{Schmittfull_1loop:2016}. The next order in perturbation theory involves two-loop integrals over two 3-D wavevectors.
While terms with at most one inverse Laplacian $\nabla^{-1}$ can again be reduced to 1-D spherical Bessel integrals, this is no longer true for more complicated terms that arise from multiple inverse Laplacians, which in Fourier space each look like $1/|\vec{k} + \vec{q}|$.
One approach \citep{Schmittfull_2loop:2016} to evaluate these integrals is to express them as
\begin{align}
  \label{eq:4}
  \Xi^{(i+j)}_{nn'}(\vr,\vk) & \equiv  \int\frac{d^3\vec{q}}{(2\pi)^3}\frac{e^{i\vq\cdot\vr}}{q^{2n}|\vk+\vq|^{2n'}}\,P^i(q)P^j(|\vk+\vq|)
\end{align}
where $i,j$ take on the values $0$ or $1$ and $P$ is the linear power spectrum.
These fundamental integrals can be reduced by expanding in Legendre polynomials in $\hat{r}\cdot\hat{k}$. The coefficients in this expansion are then
\begin{align}
  \label{eq:QNIntegral}
Q^{(N)}_{nn',L}(r,k)
= &\,
4\pi\int ds\;s^2 j_L(ks)
R^{(N)}_{n'}(s)
\int \frac{dq}{2\pi^2}\,
\non\\
&\times
q^{2-2n}
j_L(qs)j_L(qr)
\tilde R^{(N)}(q),
\end{align}
where $R$ and $\tilde R$ are defined in \cite{Schmittfull_2loop:2016}.
The  inner integral over $q$ involves two spherical Bessel functions. Thus the acceleration scheme for these integrals presented in the current work would accelerate evaluating the 2-loop SPT power spectrum.\footnote{\cite{Schmittfull_2loop:2016} also presents an alternative expansion of these integrals as an infinite sum of 1-D integrals that can be evaluated using FFTs (their equation 51). This latter expansion is actually used to evaluate their example integrals in their Figure 2, since in practice the infinite sum is relatively well-converged after a modest number of terms.}

Recent work \citep{Slepian_PT:2018} shows how these 2-loop PT integrals may be evaluated as a sum of serial back and forth 1-D FFTs. This work also uses decomposition into $f$-tensors. While the transforms there are already more or less 1-D, the 1-D transforms are done over a number of discrete integer arguments for second sBFs in each integration. Treating these integer arguments as additional free variables thus renders the 1-D transforms 2-D, and they could then be accelerated with the ideas presented in this work. This might ultimately be faster than doing serial 1-D transforms at many fixed integer values of the argument of the second sBF.

\subsection{Cyclically Summing the Perturbation Theory 3PCF Model}
The tree-level 3PCF in Standard Perturbation Theory involves two linear density fields and a second-order density field, and its calculation can be simplified by assuming that the second order density field sits at the origin of coordinates. A multipole expansion of the 3PCF can then be made by projecting its triangle-opening-angle dependence onto Legendre polynomials of the cosine of the angle enclosed by two sides extending from the origin, only $\ell = 0,1$, and $2$ modes are required to describe the opening-angle-dependence (\citealt{Szapudi2004}, \citealt{SE_RV_sig}).

However, the 3PCF we then observe in this basis does not necessarily place the second-order density field at the origin from which the triangle opening angle is also measured. Thus to connect the PT model with the observable 3PCF we must cyclically sum over the three possible locations for the second-order density field.  Two terms in the cyclic sum will be the same because once the second-order field is moved away from the origin, there is a labeling switch symmetry.

A naive way to obtain the full cyclic sum's multipole expansion is simply to solve directly for the angles and side lengths that enter and then reproject onto Legendre polynomials. However, as shown in \cite{RSD_model} \S3, the cyclic sum's multipole moments may be more efficiently obtained if all of the angular integrations are performed analytically instead. That work found that the additional moments generated by cyclic summing take the form
\begin{align}
\xi^{[L]}(r_1)\xi^{[L]}(r_3)\mathcal{L}_L(\hat{r}_1\cdot\hat{r}_3)= \sum_\ell \zeta_\ell(r_1, r_2)\mathcal{L}_\ell(\hat{r}_1\cdot\hat{r}_2).
\end{align}
$r_1$ and $r_2$ enclose the angle with respect to which we measure the 3PCF multipole coefficients $\zeta_{\ell}$, and the lefthand side is the simple-looking term that appears in the cyclic sum but which actually needs to be reprojected onto multipoles with respect to $\hat{r}_1 \cdot \hat{r}_2$.

We have defined
\begin{align}
&\zeta_l(r_1, r_2) =\frac{4\pi}{2L+1} \xi^{[L]}(r_1)\sum_{L_1}i^{l+L-L_1}\mathcal{C}^2_{L l L_1}\left(\begin{array}{ccc}
L & l & L_1\\
0 & 0 & 0
\end{array}\right)^2\nonumber\\
&\times\int\frac{k^2 dk}{2\pi^2} P(k)j_{L_1}(kr_1)j_l(kr_2),
\label{eqn:3pcf_cyc_sum}
\end{align}
where the $2\times 3$ matrix is a Wigner 3-j symbol (\citealt{AWH13}, \citealt{VMK}),
\begin{align}
\mathcal{C}_{L l L_1} =\sqrt{\frac{(2L+1)(2l+1)(2L_1+1)}{4\pi}}
\label{eqn:script_C}
\end{align}
and
\begin{align}
\xi_L(r) \equiv \int \frac{k^2 dk}{2\pi^2}j_L(kr) P(k).
\label{eqn:xi_l_defn}
\end{align}
Thus we see that accelerating the computation of the integral  on the righthand side of equation (\ref{eqn:3pcf_cyc_sum}) would render cyclically summing the PT prediction for the 3PCF more efficient. This in turn would be useful in the context of embedding a 3PCF analysis in an MCMC parameter estimation routine where model predictions must be computed many times as the cosmology is varied.

\subsection{Evaluating the Gaussian Random Field Covariance of the 3PCF}
Another context in which the integrals discussed in this work appear is the covariance matrix of the 3PCF. In particular, the leading-order covariance of the large-scale 3PCF stems from the Gaussian Random Field (GRF) component of the 6-point function, i.e. the disconnected piece. This point is also true for the 2PCF covariance, but we defer discussion of that to \S\ref{sec:Covar}. This holds both for the covariance of the isotropic 3PCF and that of the anisotropic 3PCF. This latter generalizes the isotropic covariance by promoting the isotropic power spectrum to a multipole series depending on the angle to the line of sight of the different wave vectors entering the calculation.

As shown in \cite{SE_3pt_alg}, the isotropic 3PCF covariance between 3PCF multipoles $\ell$ and $\ell'$ and triangle sides, respectively, $r_1,r_2$ and $r_1', r_2'$, takes the form
\begin{align}
&{\rm Cov}_{ll'}(r_{1},r_{2};r_{1}',r_{2}')=\frac{4\pi}{V}(2l+1)(2l'+1)(-1)^{l+l'}\nonumber\\
&\times\int r^{2}dr\sum_{l_{2}}(2l_{2}+1)\left(\begin{array}{ccc}
l & l' & l_{2}\\
0 & 0 & 0
\end{array}\right)^2\nonumber\\
&\times\bigg\{(-1)^{l_2}\xi_{0}(r)\bigg[f_{l_{2}ll'}(r;r_{1},r_{1}')f_{l_{2}ll'}(r;r_{2},r_{2}')\nonumber\\
&+f_{l_{2}ll'}(r;r_{2},r_{1}')f_{l_{2}ll'}(r;r_{1},r_{2}')\bigg]+(-1)^{(l+l'+l_{2})/2}\nonumber\\
&\times\bigg[f_{ll}(r;r_{1})f_{l'l'}(r;r_{1}')f_{l_{2}ll'}(r;r_{2},r_{2}')\nonumber\\
&+f_{ll}(r;r_{1})f_{l'l'}(r;r_{2}')f_{l_{2}ll'}(r;r_{2},r_{1}')\nonumber\\
&+f_{ll}(r;r_{2})f_{l'l'}(r;r_{1}')f_{l_{2}ll'}(r;r_{1},r_{2}')\nonumber\\
&+f_{ll}(r;r_{2})f_{l'l'}(r;r_{2}')f_{l_{2}ll'}(r;r_{1},r_{1}')\bigg]\bigg\}.
\label{eqn:fullcovar_final}
\end{align}
$\xi_0$ is defined by setting $L=0$ in equation (\ref{eqn:xi_l_defn}). We note that $l_2$ is bounded by triangle inequalities once $l$ and $l'$ are fixed. The anisotropic covariance, between 3PCF coefficients at momenta $l$ and $l'$ and shared spin $m$, with triangle sides as in the isotropic case, is
\begin{align}
&{\rm Cov}_{l_1 l_2m,l_1' l_2'm'}(r_1, r_2;r_1', r_2')=\frac{(4\pi)^{3/2}}{V}(-1)^{m +m'}(-i)^{l_1 + l_2 +l_1' + l_2'} \nonumber\\
&\times \int r^2 dr\sum_{l_q l_p l_k} \frac{1} {\sqrt{(2l_q + 1)(2l_p + 1)(2l_k +1)}}\nonumber\\
&\times \sum_{J_1 J_2 J_3} \mathcal{D}_{J_1 J_2 J_3}\mathcal{C}_{J_1 J_2 J_3}\left(\begin{array}{ccc}
J_1 & J_2 & J_3\\
0 & 0 & 0
\end{array}\right)\nonumber\\
&\times \Bigg\{ \xi_{l_k}(r)\bigg[ w_1 f^{l_q}_{J_1 l_1 l_1'}(r; r_1, r_1') f^{l_p}_{J_2 l_2 l_2'}(r; r_2, r_2') \nonumber\\
&+ w_2 f^{l_q}_{J_1 l_1 l_2'}(r; r_1, r_2') f^{l_p}_{J_2 l_2 l_1'}(r; r_2, r_1')\bigg] +
\left(\begin{array}{ccc}
J_1 & J_2 & J_3\\
S_1 & S_2 & S_3
\end{array}\right)\nonumber\\
&\times \bigg\{ f^{l_q}_{J_1 l_1}(r; r_1) \bigg[w_3 f^{l_p}_{J_2 l_2 l_2'}(r; r_2, r_2') f^{l_k}_{J_3 l_1'}(r; r_1')\delta^K_{S_1 -m, S_3 -m'} \nonumber\\
&+ w_4 f^{l_p}_{J_2 l_2 l_1'}(r; r_2, r_1') f^{l_k}_{J_3 l_2'}(r; r_2')\delta^K_{S_1 -m, S_3 m'} \bigg]\nonumber\\
&+  f^{l_p}_{J_2 l_2}(r; r_2) \bigg[w_5 f^{l_q}_{J_1 l_1 l_2'}(r; r_1, r_2') f^{l_k}_{J_3 l_1'}(r; r_1')\delta^K_{S_2 m, S_3 -m'} \nonumber\\
&+ w_6 f^{l_q}_{J_1 l_1 l_1'}(r; r_1, r_1') f^{l_k}_{J_3 l_2'}(r; r_2')\delta^K_{S_2 m, S_3 m'} \bigg] \bigg\}\Bigg\}.
\label{eqn:covar_w_wts}
\end{align}
$\mathcal{C}$ is defined in equation (\ref{eqn:script_C}), $\xi_{l_k}$ is defined in equation (\ref{eqn:xi_l_defn}), $\mathcal{D}_{J_1 J_2 J_3}\equiv i^{J_1 + J_2 + J_3}$, $\delta^{\rm K}$ is a Kronecker delta, unity if its arguments are equal and zero otherwise. The $w_i$ are angular momentum coupling weights involving integrals of three and four spherical harmonics, i.e. Gaunt integrals and products of Gaunt integrals, or equivalently, Wigner 3j-symbols.

The main cost of evaluating these covariance expressions is obtaining the $f$-tensors (defined in equation \ref{eqn:f_tensor}; see also equation 45 of \citealt{Slepian:2018}), and again one might wish to compute them many times if the 3PCF were used as part of an MCMC cosmological parameter analysis. A fast means of computing them is thus highly enabling for a 3PCF analysis where cosmological parameters are varied.

\subsection{Gaussian Random Field Contribution to N-Point Correlation Functions}
In the future it is likely that we will measure higher-point statistics of galaxy clustering. All of the even correlation functions (e.g. 4PCF, 6PCF, 8PCF) will have contributions from the Gaussian Random Field component of the overall density field. However, this component simply duplicates information already available in the 2PCF. Thus removing it before fitting models to the higher-order (even) correlation functions is desirable. It must therefore be calculated.

Several recent works have discussed measurement of higher $N$-point functions. \cite{Sabiu:2019} uses a graph spatial database to obtain NPCFs, and presents the first measurement of the 4PCF of Luminous Red Galaxies in the Sloan Digital Sky Survey Baryon Oscillation Spectroscopic Survey (SDSS BOSS) CMASS sample. \cite{Tomlinson:2019} outlines how to measure higher-order polyspectra, the Fourier-space analogs of the NPCFs, averaging over internal angles of the Fourier-space polyhedron. Slepian, Cahn \& Eisenstein (in prep.) outlines how to measure higher-order correlation functions in the basis of rotation-invariant combinations of spherical harmonic coefficients, building on \cite{SE_3pt_alg}, \cite{SE_3pt_FT}, \cite{SE_aniso_3pt}. This will then require removing remove the contributions of the GRF piece in the harmonic basis. Here we sketch results in configuration space as these give rise to the double-Bessel transforms of the power spectrum that are our focus in this work.

For an even N-point function, the Gaussian random field contribution will be the cyclic sum of products of 2-point functions with all possible vector argument differences (shown in Slepian, Cahn \& Eisenstein in prep.). For instance, for the 4PCF, denoted $\zeta^{(4)}$, we have
\begin{align}
\zeta^{(4)}_{\rm GRF}(\vec{r}_1, \vec{r}_2, \vec{r}_3, \vec{r}_4) = \xi(|\vec{r}_2 -\vec{r}_1|) \xi(|\vec{r}_4 -\vec{r}_3|)  + {\rm cyc.}
\end{align}
For the 6PCF, we have
\begin{align}
&\zeta^{(6)}_{\rm GRF}(\vec{r}_1, \vec{r}_2, \vec{r}_3, \vec{r}_4, \vec{r}_5, \vec{r}_6) =\nonumber\\
& \xi(|\vec{r}_2 -\vec{r}_1|) \xi(|\vec{r}_4 -\vec{r}_3|) \xi(|\vec{r}_6 -\vec{r}_5|)  + {\rm cyc.}
\end{align}
To compute the GRF piece of an even NPCF projected onto the basis of spherical harmonics, we may do so by writing each correlation function as an inverse Fourier Transform of the power spectrum, and then projecting the plane waves therein onto spherical harmonics using the plane wave expansion and then orthogonality. This will leave $f$-tensors of exactly the form given in equation (\ref{eqn:f_tensor}). Note that we may indeed have $\ell \neq \ell'$. The only constraint on each term in the cyclic sum for an even NPCF is that the parity of the summed angular momenta in a given term be even. This is required by rotation-invariance: each spherical harmonic picks up $(-1)^\ell$ as it rotates, we need the sum of the $\ell$ to be even.

We also note that in analysing the results of an NPCF computation, covariance matrices will be required for optimally weighting the data. These quickly become enormous matrices and for generic NPCFs would require a prohibitive number of mock catalogs if one wished to estimate them from mocks. Instead, \cite{Xu:2013}, \cite{SE_3pt_alg}, \cite{SE_RV_constraint}, \cite{SE_aniso_3pt}, and Slepian, Cahn, \& Eisenstein (in prep.) advocate using a template covariance matrix; the last work referenced extends this to general NPCFs and shows that the template can be cast solely in terms of the integrals (\ref{eqn:f_tensor}). Thus speeding up the evaluation of these will aid computation of the NPCF covariance matrix templates.

\section{Main Example: Gaussian Random Field 2PCF Covariance Matrix}
\label{sec:Covar}
The main numerical example we will present in this work is the covariance matrix of the anisotropic 2PCF multipoles. It can be shown (e.g. \citealt{Xu:2013}) that this matrix is given by
\begin{align}
{\rm Cov}\left(\xi_{\ell}(r),\xi_{\ell'}(r')\right) =\int \frac{k^2dk}{2\pi^2}\;j_{\ell}(kr) j_{\ell'}(kr')P^2(k)
\label{eqn:2pcf_covar}
\end{align}
with $P$ the linear power spectrum. If one wishes to include shot noise one can promote $P$ to $P+1/n$ with $n$ the survey number density, and often one also uses a ``template'' power spectrum where the BAO features have been smoothed, usually by a Gaussian kernel of width $\sim 7-8\;{\rm Mpc}$, which is the root-mean-square displacement in the Zel'dovich approximation. One also generally smoothes the whole power spectrum by a Gaussian of width $1\;{\rm Mpc}$ to avoid ringing due to the truncation of the Fourier-space grid. Generally the radial variables in the covariance are binned to match the fact that the 2PCF itself is measured on radial bins.

Here we will simply do the full, unbinned calculation, since certainly if that is accurate, the binned version will end up more so, as binning effectively smoothes the spherical Bessel functions, making the numerical accuracy requirements for their integration less stringent. One could also imagine replacing the spherical Bessel functions with their binned analogs before performing the integration over $k$, as for instance \cite{Xu:2013} does (their equations 8-11). That would still be amenable to the techniques outlined in this work, since due to their recursion relations, the integral of a spherical Bessel functions yields a sum of two others, weighted by powers of the argument (e.g. \cite{NIST_DLMF} 10.51.1 and 10.51.2 can be manipulated to show this). Consequently even after ``analytic'' binning one still has integrals over $k$ of the type we focus on performing here.

The monopole and quadrupole have been measured by BOSS, and DESI will measure them to high precision. DESI will not achieve very high precision ($\sim 10\%$) on the hexadecapole $\ell=4$  \citep{DESI:2016}, so we further focus our numerical work here down to the monopole $(\ell = 0)$ and quadrupole $(\ell = 2)$ and their cross-covariances as well.

In particular, the use case numerical-results we present (i.e. those where we have used the true power spectrum, rather than some test power spectrum for simply verifying the accuracy of our integrations) will be doing the integrals for ${\rm Cov}(\zeta_0,\zeta_0)$ and ${\rm Cov}(\zeta_2,\zeta_2)$. We do not perform tests for ${\rm Cov}(\zeta_2,\zeta_0)$ because if we are able to get the other two matrix elements right, we expect that the numerical performance will be sufficient for this cross term.

\section{Rotation Method: From 2-D to 1-D}
\label{sec:technique}
\subsection{Basic Idea}
\label{subsec:basic}
We now show how to reduce $f_{\ell \ell'}$ (equation \ref{eqn:f_tensor}) to a sum of 1-D integrals. All
spherical Bessel functions $j_{\ell}(x)$ are of the form
\begin{align}
j_{\ell}(x)=\frac{C_{\ell}(x)}{x^{\ell}}\cos x+\frac{S_{\ell}(x)}{x^{\ell+1}}\sin x,
\end{align}
where $C_{\ell}(x)$ and $S_{\ell}(x)$ are polynomials with a finite, small
number of terms (``C'' denotes the polynomial multiplying the cosine, ``S'' that multiplying the sine). Writing further that
\begin{align}
C_{\ell}(x)=\sum_{n=0}^{n_{{\rm max}}}c_{\ell n}x^{n}
\end{align}
and
\begin{align}
S_{\ell}(x)=\sum_{m=0}^{m_{{\rm max}}}s_{\ell m}x^{m},
\end{align}
where $n_{{\rm max}}\leq \ell-1$ for odd $\ell$ and $n_{{\rm max}}\leq \ell-2$
for even $\ell$ and $m_{{\rm max}}\leq \ell - 1$ for odd $\ell$ and $m_{{\rm max}}\leq \ell$
for even $\ell$. Exact forms for these relations are presented in Appendix \ref{app:ExplicitJlJlExpansion}. We then see that a product of two spherical Bessel
functions with arguments $ka$ and $kb$ will have the form
\begin{align}
\label{eqn:fundam}
&j_{\ell}(ka)j_{\ell'}(kb)=\\
&\cos ka\cos kb\left(k^{-\ell-\ell'}a^{-\ell}b^{-\ell'}\right)\sum_{nn'}c_{\ell n}c_{\ell'n'}k^{n+n'}a^{n}b^{n'}\nonumber\\
&+\cos ka\sin kb\left(k^{-\ell-\ell'-1}a^{-\ell}b^{-\ell'-1}\right)\sum_{nm'}c_{\ell n}s_{\ell'm'}k^{n+m'}a^{n}b^{m'}\nonumber\\
&+\sin ka\cos kb\left(k^{-\ell-1-\ell'}a^{-\ell-1}b^{-\ell'}\right)\sum_{mn'}s_{\ell m}c_{\ell'n'}k^{m+n'}a^{m}b^{n'}\nonumber\\
&+\sin ka\sin kb\left(k^{-\ell-\ell'-2}a^{-\ell-1}b^{-\ell'-1}\right)\sum_{mm'}s_{\ell m}s_{\ell'm'}k^{m+m'}a^{m}b^{m'}.\nonumber
\end{align}
$c_{\ell n},\;c_{\ell'n'}$, $s_{\ell m}$, and $s_{\ell' m'}$ are constants; the argument $b$ always matches to primed indices. It is clear that the powers of $a$ and $b$ can be taken outside
any integrals over $k$. We note that since all spherical Bessel functions
have asymptotic behavior $j_{\ell}(x)\to 1/x$ for large $x,$ their product
as above must fall as $1/k^{2}$. This can be directly verified case
by case (product of even and odd, two even, or two odd spherical Bessel
functions) and term-by-term above using the maximal values of $n,m,n'$
and $ $$m'$ noted. Importantly, this holds individually for each
of the four terms in the overall sum above. This large-$k$ behavior
cancels the $k^{2}$ in the integration measure, and assures that
the integrals $f_{\ell \ell'}$ will not be UV-divergent as long as $P(k)\to0$
as $k\to\infty.$

For small values of their argument, $j_{\ell}(x)\to x^{\ell}$ so their
product as above scales as $ $$k^{\ell+\ell'}$. As long as the power spectrum
satisfies $P(k)\propto k^{n}$ as $k \to 0$ with $n \geq -(\ell + \ell' + 2)$ the
integrals $f_{\ell \ell'}$ will not be IR-divergent.

However, we note that the small-argument scaling of the spherical
Bessel functions relies on cancellation of terms in a Taylor series,
and that the splitting above, into an overall sum over four terms,
does not retain this cancellation on a term-by-term basis. For instance,
the last term above will scale as $k^{-\ell-\ell'}$ as $k\to 0.$ Consequently,
if each of the four terms above is integrated individually against
the power spectrum, we are not guaranteed to avoid IR-divergence.
However, numerical integration over a finite range effectively regularizes this divergence,
as we discuss further in \S\ref{sec:imp}.

Overall, we see that if $\sin ka\cos kb$ and the other relevant
combinations of trigonometric functions entering can be factorized
so as to eliminate the $a$ and $b$ dependence from the $k$ integral,
we will have fully reduced the seemingly 2-D problem of computing
$f_{\ell \ell'}$ to a sum over 1-D integrals.

Defining $\Delta = a - b$ and $\Sigma = a + b$ and employing the identities
\begin{align}
&\cos ka\cos kb =\frac{1}{2}\left[\cos k\Delta +\cos k\Sigma \right]\nonumber\\
&\cos ka\sin kb =\frac{1}{2}\left[\sin k\Sigma -\sin k\Delta \right]\nonumber\\
&\sin ka\sin kb =\frac{1}{2}\left[\cos k\Delta-\cos k\Sigma \right]
\end{align}
we see that each of the four terms in equation (\ref{eqn:fundam}) will lead to a
sum over integrals that each look like
\begin{align}
I_{C}^{q}(u)=\int\frac{k^{2}dk}{2\pi^{2}}P(k)k^{q}\cos ku
\label{eqn:cosine_integral}
\end{align}
and
\begin{align}
I_{S}^{p}(u)=\int\frac{k^{2}dk}{2\pi^{2}}P(k)k^{p}\sin ku
\label{eqn:sine_integral}
\end{align}
with $u \to \Sigma$ or $\Delta$. $I$ is for ``integral'', $C$ for ``cosine'', $S$ for ``sine'', and the superscript $p$ indicates the power of $k$ weighting the integrand.

\begin{figure}
    \centering
    \includegraphics[width=\linewidth]{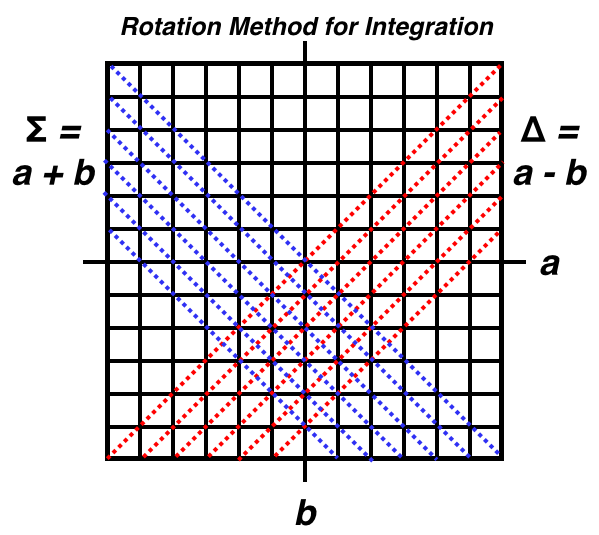}
    \caption{Here we illustrate how the ``rotation'' method works, for the case of a two-sBF integral. $\Sigma$ and $\Delta$ correspond to rotations of 45 degrees, which essentially decouple the 2-D integrals over $(a,b)$ into sums and differences of 1-D integrals over $\Delta$ and $\Sigma$. This decoupling is analogous to diagonalizing a matrix.
    }
    \label{fig:rotation_method}
\end{figure}

From these variables we can see that writing the products of trigonometric
functions as a sum of trigonometric functions effectively performed
a 2-D, 45 degree rotation from the $a, b$ to the $\Sigma, \Delta$ plane. This is shown schematically in Figure \ref{fig:rotation_method}.

These integrals are simply the Fourier cosine and sine transforms
of $P(k)k^{2+q}$ and $ $$P(k)k^{2+p}$ respectively, and are easily
evaluated by a standard FFT package (for a linearly-spaced grid) or FFTLog package (for a logarithmically-spaced grid; e.g. \citealt{Hamilton:2000}).

\subsection{Examples and Automatic Generation of the Required Forms}
The complexity required to make our notation general
may obscure that for reasonably small $\ell$ and $\ell'$, explicit computation
shows that the required number of terms is small. For instance,
for $\ell = 0 = \ell',$ we have
\begin{equation}
j_{0}(ka)j_{0}(kb)=\frac{1}{2k^{2}ab}\left\{ \cos k(a-b)-\cos k(a+b)\right\}
\end{equation}
so that
\begin{equation}
f_{00}(a,b)=\frac{1}{ab}\left\{ I_{C}^{-2}(\Delta)-I_{C}^{-2}(\Sigma)\right\}.
\end{equation}
Only one fundamental integral is required for this simple case above, as the integral once obtained may be evaluated at the required arguments $\Sigma$ and $\Delta$ as long as one has computed it over a large enough range. Furthermore, we emphasize that if one uses the same grid in $a$ and
$b$, of length $N_{a},$ there are not $N_{a}^{2}$ unique values
of $a \pm b,$ but rather $3N_{a}$; we can subtract all elements
$b$ from the $a = 0$ element, generating a negative $a - b$ grid
of length $N_{a}$, and we can add all elements of the $b$ grid
to the maximal $a$ element, generating an additional positive grid
from $a_{{\rm max}}$ to $2a_{{\rm max}}$. But we need not evaluate
the integrals at all $3N_{a}$ of these points; rather one can use
the parity properties of the Fourier sine and cosine transform
to avoid evaluating on the negative portion of the net grid, reducing
the work to $2N_{a}$. In our numerical work we did not implement this latter trick to avoid the need for carefully tracking signs, but adding it would doubtless further accelerate the calculation, likely by roughly 30\% as it takes the work from $3N_a$ to $2N_a$.

For $\ell = 1 = \ell'$, we have
\begin{align}
&f_{11}(a, b) = \frac{1}{2ab}\bigg\{\frac{1}{ab}I_{C}^{-4}(\Delta) + I_{C}^{-2}(\Delta) - \frac{1}{a b}I_{C}^{-4}(\Sigma)
+ I_{C}^{-2}(\Sigma) \nonumber\\
&+ \frac{1}{b}I_{S}^{-3}(\Delta) - \frac{1}{a}I_{S}^{-3}(\Delta)
-  \frac{1}{b}I_{S}^{-3}(\Sigma) - \frac{1}{a}I_{S}^{-3}(\Sigma) \bigg\}.
\label{eqn:j1j1_expansion}
\end{align}
Here, only three fundamental integrals are required: $I_C^{-4}$, $I_C^{-2}$, and $I_S^{-3}$.
For $\ell = 2 = \ell'$ we find
\begin{align}
&f_{22}(a, b) = \frac{9}{2 a^3 b^3}\left( I_{C}^{-6}(\Delta) - I_{C}^{-6}(\Sigma)\right) -\nonumber\\
&\frac{1}{2}\left(\frac{3}{a b^3} - \frac{9}{a^2 b^2} + \frac{3}{a^3 b}\right) I_C^{-4}(\Delta) + \frac{1}{2}\left(\frac{3}{a b^3} + \frac{9}{a^2 b^2} + \frac{3}{a^3 b}\right) I_C^{-4}(\Sigma) \nonumber\\
&+ \frac{1}{2 a b}\left( I_C^{-2}(\Delta) - I_C^{-2}(\Sigma)\right) + \frac{9}{2}\left(\frac{1}{a^2 b^3} - \frac{1}{a^3 b^2}\right)I_S^{-5}(\Delta) - \nonumber\\
&\frac{9}{2}\left(\frac{1}{a^2 b^3} + \frac{1}{a^3 b^2}\right)I_S^{-5}(\Sigma) + \frac{3}{2}\left(\frac{1}{a b^2} - \frac{1}{a^2 b}\right)I_S^{-3}(\Delta)  \nonumber\\
&+ \frac{3}{2}\left(\frac{1}{a b^2} +\frac{1}{a^2 b} \right) I_S^{-3}(\Sigma).
\end{align}
Although the expression is long, only five fundamental integrals are required: $I_C^{-6}$, $I_C^{-4}$, $I_C^{-2}$, $I_S^{-5}$, and $I_S^{-3}$.

For an ``off-diagonal'' example, i.e. with $\ell \neq \ell'$, we consider
\begin{equation}
f_{10}(a, b)=\frac{1}{2a^{2}b}\left\{ I_{C}^{-3}(\Delta)-I_{C}^{-3}(\Sigma)+aI_{S}^{-2}(\Delta)-aI_{S}^{-2}(\Sigma)\right\}.
\end{equation}
In this off-diagonal example the difference in indices $(\ell = 1, \ell'=0)$ breaks the switch symmetry between $a$ and $b$, so as we might expect the final expression is not symmetric under $a \leftrightarrow b$. We also highlight that just two fundamental integrals are required.

Another example, relevant for the cross covariance of the monopole and quadrupole of the anisotropic 2PCF, is
\begin{align}
f_{02}(a,b) &= \frac{1}{2 a b} \bigg\{\frac{3}{b^2}\left( I_C^{-4}(\Delta) - I_C^{-4}(\Sigma)\right) -  \left(I_C^{-2}(\Delta) + I_C^{-2}(\Sigma)  \right)\nonumber\\
& -\frac{3}{b}\left( I_S^{-3}(\Delta) + I_S^{-3}(\Sigma) \right)\bigg\}.
\end{align}
Above only three fundamental integrals are required.

A third example is
\begin{align}
&f_{13}(a,b)=\frac{15}{2a^{2}b^{4}}\left[I_{C}^{-6}(\Delta)+I_{C}^{-6}(\Sigma)\right]\nonumber\\
&+\left(\frac{15}{2ab^{3}}-\frac{3}{a^{2}b^{2}}\right)\left[I_{C}^{-4}(\Delta)+I_{C}^{-4}(\Sigma)\right]\nonumber\\
&-\frac{1}{2ab}\left[I_{C}^{-2}(\Delta)+I_{C}^{-2}(\Sigma)\right]\nonumber\\
&+\frac{15}{2}\left(\frac{1}{ab^{4}}-\frac{1}{a^{2}b^{3}}\right)\left[I_{S}^{-5}(\Delta)+I_{S}^{-5}(\Sigma)\right]\nonumber\\
&-\left(\frac{3}{ab^{2}}-\frac{1}{2a^{2}b}\right)I_{S}^{-3}(\Delta)+\left(\frac{3}{ab^{2}}+\frac{1}{2a^{2}b}\right)I_{S}^{-3}(\Sigma).
\end{align}
Here only five fundamental integrals are required.

In general these expressions can be easily generated using \textsc{Mathematica}\footnote{We make some example \textsc{Mathematica} code available at \url{https://github.com/eelregit/sbf_rotation}}
with the command
\begin{align}
\label{eqn:mathematica_command}
&{\rm TrigReduce}[{\rm FunctionExpand}[{\rm SphericalBesselJ}[l,kr]\nonumber\\
&\times {\rm SphericalBesselJ}[l',kr']]].
\end{align}
As noted earlier, a general form for explicit expressions for the coefficients for arbitrary $\ell$ and $\ell'$ (i.e. the two-sBF case) is  given in Appendix~\ref{app:ExplicitJlJlExpansion}. We also investigated using \textsc{python}'s $sympy$ symbolic manipulation package to generate these forms. However for the relatively modest-$\ell$ cases we focus on here this was not necessary, so we defer full development of that code to future work.

\section{Generalization to three or more spherical Bessel functions}
\label{sec:general}
It is clear that all we require to generalize the approach of \S\ref{sec:technique}
to integrals of three or more spherical Bessel functions is the analogous
trigonometric identities for converting products to sums with three
or more trigonometric functions. Rewriting the trigonometric functions
as complex exponentials, multiplying them, refactoring into combinations
that depend on $a \pm b\pm c$, and resumming to trigonometric functions
yields these integrals. \textsc{Mathematica} can again be used to obtain
them quickly with a command analogous to that in equation (\ref{eqn:mathematica_command}).

We note that the number of identities needed is not large because
one can always switch $a\leftrightarrow b,$ $b \leftrightarrow c,$
etc. so that only the number of cosines and sines entering the product
matters, not which argument is in which function. For three spherical
Bessel functions, one therefore needs only four identities, and for products of four spherical Bessel functions, five identities. In general the number of combinations is simply the number of possible
factors of sine that can appear, which runs from zero up to the number
of spherical Bessel functions being considered.

The simplest example for three spherical Bessel functions is
\begin{align}
f_{000}(a, b, c) &= \frac{1}{4abc}\bigg\{-I_S^{-3}(\chi_{--}) + I_S^{-3}(\chi_{+-}) \nonumber\\
&+ I_S^{-3}(\chi_{-+}) - I_S^{-3} (\chi_{++}) \bigg\}
\label{eqn:j0j0j0_expansion}
\end{align}
with
\begin{align}
\chi_{\pm \pm} = a \pm b \pm c
\end{align}
the generalization of $\Delta$ and $\Sigma$ to three variables instead of two. The $i^{th}$ subscripted sign denotes the sign preceding the $(i + 1)^{th}$ variable.

We emphasize that the possible accelerations presented in the next section (\S\ref{sec:accel}) will
also apply to integrals against products of three or more spherical
Bessel functions, since after use of the trigonometric identities
we will again only have 1-D integrals of sine and cosine against the
power spectrum.

\section{Possible Accelerations}
\label{sec:accel}
We now present a complementary technique for the particular case of
an integral of two spherical Bessel functions against the power spectrum,
similar to the idea presented in \cite{Assassi:2017} and \cite{Gebhardt:2018}.
If the power spectrum can be expanded into a sum over (possibly complex) power laws as
\begin{align}
P(k) = \sum_n c_n k^n,
\end{align}
then we have
\begin{equation}
f_{\ell \ell'}(a,b)=\sum_{n}c_{n}\int\frac{k^{2}dk}{2\pi^{2}}k^{n}j_{\ell}(ka)j_{\ell'}(kb).
\label{eqn:power_spec_complex}
\end{equation}
Using the identity that
\begin{equation}
j_{\ell}(x)=\sqrt{\frac{\pi}{2x}}J_{\ell+1/2}(x),
\end{equation}
where $J_{\ell+1/2}$ is a Bessel function, we find
\begin{equation}
f_{\ell \ell'}(a, b)=\frac{1}{4\pi\sqrt{ab}}\sum_{n}c_{n}\int kdk\; k^{n}J_{\ell+1/2}(ka)J_{\ell'+1/2}(kb).
\label{eqn:GR_integral_one}
\end{equation}
Using \cite{Gradshteyn:2007} 6.574.1-3, the integral (\ref{eqn:GR_integral_one}) can be
evaluated explicitly in terms of Gamma functions $\Gamma$ and hypergeometric
functions $F.$ We have the restrictions that $\ell + \ell'+2>-n$ and $n<1.$
The argument of the hypergeometric function depends on the ratio $(a/b)$
or $(b/a)$, whichever is less than unity (6.574.1 for $a<b$, 6.574.3
for $b<a,$ 6.574.2 for $a = b$). For grids in $a$ and $b$ of
length $N_{a},$ we will thus require $N_{a}^{2}$ evaluations of
the hypergeometric function. However fast implementations of these functions
exist and this should not be a significant computational burden. This approach was used for the case where $\ell = \ell'$ in \cite{Assassi:2017}, with further discussion in \cite{Gebhardt:2018}. The latter focuses on a stable, accurate, and fast method for evaluating the required hypergeometric functions.

We can use this approach to evaluate an integral over three spherical
Bessel functions as well, a problem not considered in previous literature. With the power spectrum again expanded as in equation (\ref{eqn:power_spec_complex}) we have
\begin{align}
&f_{\ell \ell' \ell''}(a, b, c)=\frac{1}{2^{5/2} \pi^{1/2} abc}\nonumber\\
&\times \sum_{n}c_{n}\int kdk\; k^{n}J_{\ell+1/2}(ka)J_{\ell'+1/2}(kb)J_{\ell''+1/2}(kc)
\label{eqn:3_bessel_overlap}
\end{align}
which can be evaluated in terms of an Appell hypergeometric function using \cite{Gradshteyn:2007} 6.578.1.

In general, the approach we have outlined above can be extended to obtain integral-free
forms for an integral of the power spectrum against an arbitrary number
of spherical Bessel functions. This can be done using the technique outlined in \cite{Fabrikant:2003} and \cite{Fabrikant:2013} for computing integrals of an arbitrary number of spherical
Bessel functions against power laws. For instance \cite{Fabrikant:2013} equations (16) and (19) can
be combined to provide the result for the case of four spherical Bessel
functions. \cite{Fonseca:2017} provides a publicly-available \textsc{Mathematica} notebook implementing Fabrikant's procedure for three spherical Bessel functions.\footnote{\url{https://zenodo.org/record/495795},\\ \textsc{FabrikantIntegrals.nb}.}

\section{Numerical Work}
\label{sec:imp}
Here we briefly outline our numerical tests of these ideas. Along with this paper we make publicly available a simple \textsc{python} code implementing the rotation method (but with direct integration, not FFTLog) and the naive method for comparison.\footnote{All code for this paper is available at \url{https://github.com/eelregit/sbf_rotation}} This version uses only simple functions such as sine and cosine and Romberg integration, so it is backwards-compatible with all \textsc{python} versions and might be well-suited for integration into a larger package that focuses on back-compatibility, such as \textsc{COLOSSUS} (\citealt{Diemer:2018}). We also make available through \texttt{mcfit} an implementation of the rotation method that further exploits FFTLog to render the integrals even faster, as discussed at the end of \S\ref{subsec:basic}.\footnote{\texttt{mcfit} is a broader custom \textsc{python} package developed by one of us (YL) to do a number of common problems in cosmology. It is available with documentation at \url{https://github.com/eelregit/mcfit}} We recommend this latter be used for ``production'' analyses, with the former mainly of use for understanding the ideas here as well as offering a simple implementation that requires no external libraries or packages save Romberg integration.

For most of our numerical work, we focused on the cases $j_0(ka) j_0(kb)$ (``00''), $j_1(ka) j_1(kb)$ (``11''), $j_2(ka) j_2(kb)$ (``22''), and $j_0(ka) j_0(kb) j_0(kc)$ (``000''), on grids where $a$, $b$, and if relevant $c$ were linearly spaced from $0-100$ in steps of unity. The $00$ and $22$ cases are relevant for the anisotropic 2PCF covariance matrix. We did the $11$ case for completeness and the $000$ case to show how our method extends to triple-sBF integrals. The anisotropic 2PCF covariance matrix can also have cross terms (e.g. 02); we do not investigate these numerically because the issues should be just the same as for the auto terms. In particular, 02 will have less stringent numerical requirements than 22 because the divergences we have to track go up with higher total $\ell$. Finally, we performed tests using a case where the answer is exactly known analytically over a much larger range in $a$ and $b$, from $10^{-4}$ to $10^{2}$ in each.

\subsection{Overview}
We compared three approaches. For this section only, to make our plot titles complete yet concise, we adopt a shorter notation for our integrals than the $f$-tensor notation of earlier sections. We only consider a weight of $k^2$, so we do not need the upper index of the $f$ tensors. However, we consider integrals of both $P$ and $P^2$, so we need an index for that as well as two or three for the orders of each sBF integrated over. Finally, we suppress the arguments of the integral since that will be evident from the plot axis titles. We thus use the notation
\begin{align}
I_{\ell \ell'}^{[n]}\equiv \int \frac{k^2 dk}{2\pi^2}\;j_{\ell}(ka) j_{\ell'}(kb) P^{n}(k).
\end{align}
We will also always weight our results by $(ab/100^2)^2$ as this is the weight that each pixel of the result would have on a spherical shell; this is relevant if one were to further integrate these results onto spherical bins.
\subsubsection{Naive approach: $\Ng^3$}
\label{subsubsec:ncubed}
First, we evaluated these integrals using a naive approach scaling as $N_a N_b N_k$. Here, $N_a = N_b$ is the number of grid points in $a$ and $b$, and $N_k$ is the number of points in $k$ used for the integration. The scaling is $N_a N_b N_k$ because at each point in the $(a,b)$ plane, we need to evaluate the integrand at each of the $N_k$ points in $k$ and then sum (or use some other integration scheme of choice). As a shorthand we will sometimes refer to this as the $\Ng^3$ approach, which would be the case if one used equal grids in $a$, $b$, and $k$ of length $\Ng$ each. In actual practice, for us $N_k \gg N_a$.

For the case of $j_0(ka)j_0(kb)$, the result was not sensitive to whether we used e.g. a Romberg integration method or a simple sum over the sample points. For the other cases (higher-$\ell$), we used Romberg as it seemed a simple sum was not sufficiently precise. Furthermore, we examined both fixed-grid and adaptive Romberg. The fixed grid tests were all on a linearly-spaced grid of $4097$ points from $k = 2\times 10^{-4}$ to $2\;h$/Mpc. We used $4097$ points because the \textsc{Python} (\texttt{scipy}) implementation of Romberg integration we used required a grid of length $2^p + 1$, with $p$ a natural number. For the adaptive Romberg tests we explored several different maximal $k$; we discuss this in greater detail for in the sections on the specific tests where it became relevant.

\subsubsection{Rotation method with direct integration: $\Ng^2$}
\label{subsubsec:nsquared}
Our second implementation was the rotation method but with direct integration. It scales as $N_a N_k$, and as a shorthand we will sometimes refer to it as the $\Ng^2$ approach, which would be the case if one used equal grids in $a$, $b$, and $k$. In this second implementation, we simply employed Romberg integration to obtain each of the required 2-D integrals from $k$ to all $\Delta$ or all $\Sigma$ (e.g. equations \ref{eqn:cosine_integral} and \ref{eqn:sine_integral}). This approach scales as $N_{\Delta}N_k$ (or $N_{\Sigma}N_k$), and $N_{\Delta} = N_{\Sigma} = 2N_a$. We explored Romberg both on a fixed $a$ $priori$ grid and with adaptive sampling. The latter required defining an interpolating function for the integrand and allowing the integration routine to call it. This latter approach is slower than simply using a fixed grid, as each call to the interpolant adds some time. However the adaptive method is more accurate. The fixed-grid Romberg was again on a linearly-spaced grid with the same parameters as in \S\ref{subsubsec:ncubed}.

We compared adaptive-to-adaptive for the naive and rotation methods, and within the ``adaptive'' tests we also explored how much the maximal number of sub-divisions we permit the Romberg algorithm to make impacts the comparison.

To attempt to reduce the dynamic range of the integrals before adding them up and canceling regularized-divergent terms,\footnote{Some of the integrals do in fact diverge if taken from $k=0\to\infty$, but of course numerically integrating them over a finite range regularizes this divergence.} we also explored integrating on shells in $k$, i.e. evaluating each 1-D integral on a set of bins in $k$. We could then add up all of the 1-D integrals on each bin, and then finally add over bins. The aim of this scheme was to reduce the dynamic range one must track before getting to cancellations. We hoped that the cancellations on a given bin would place less stringent demands on the numerical precision. This point probably does hold true, but unfortunately an integral that is performed by being split into bins, and then doing a Romberg integration on each bin, is no longer globally Romberg. This is because Romberg weights depend on the whole range being integrated over. Consequently the Romberg weights set up over just one bin do not match the Romberg weights one would use on integration points in that bin if one were performing a Romberg integration on the whole range covered by all the bins. The loss of precision from failure to remain globally Romberg turned out to outweigh the possible gain due to performing the cancellations bin by bin. We therefore summarize this line of inquiry as a cautionary note only, but do not show detailed results from it.

\subsubsection{Rotation method with FFTLog-integration}
Finally, we produced an implementation of the rotation method using FFTLog (e.g. \citealt{Hamilton:2000}) to obtain the integrals we required. This approach scales as $N_k \log N_k$ (and requires that $N_k = N_{\Sigma} = N_{\Delta}$), and so is more efficient than the rotation method implementation with direct integration. It also has the advantage that it naturally handles logarithmically-spaced inputs, allowing much more dynamic range in e.g. the power spectrum, and matching the typical output format of linear Boltzmann solvers such as CAMB (\citealt{Lewis:2000}).

\subsection{Brief Summary}
We now briefly summarize our numerical test results, focusing on precision and timing. For readers who simply want to apply our method to the low-$\ell$ cases we have worked out, this subsection should be enough. For readers interested in the details of our numerical tests, two subsections following this one provide more extensive discussion (\S\ref{subsec:unsmoothed} and \S\ref{subsec:squared}).

\subsubsection{Precision}
\label{subsubsec:precision}
Overall, we were able to obtain percent to sub-percent agreement between the results of the 3 methods. For most applications this should be sufficient. Doubtless for further, application-dependent precision needs, one could refine the details of the integration approaches and parameters, e.g. using smoothing, using a finer $k$-grid, requiring higher precision in the integration, using a different $k$-range, using different numerical types. For our $numerical$ test cases (power spectrum, power spectrum squared), we caution that if one wished to push beyond the precision found here, it would as a first step be worth checking that the fiducial case of the naive, $\Ng^3$ implementation even is itself accurate to that level. Of course this question is well-posed in our analytic test case, where the exact answer is known. Finally, testing the $\Ng^2$ method against the $\Ng \log \Ng$ method required some iteration to sufficiently match the dynamic range accessible to the latter with the former. We found that $k_{\rm max} \simeq 8\;h$/Mpc was enough to obtain very good agreement over the range of $a$ and $b$ investigated for $P$, and $k_{\rm max} \sim 4\;h$/Mpc for $P^2$. As further detailed in \S\ref{subsec:squared}, going to higher $k_{\rm max}$ in the $\Ng^3$ or $\Ng^2$ methods actually hurt the agreement with the $\Ng \log \Ng$ method, likely due to details of how \textsc{python}'s Romberg integration routine handles the high-$k$ oscillations of the sines and cosines.

We further note that the areas of largest fractional disagreement between methods tended to be those where the integrals had smallest absolute value. It is very easy to get a numerical calculation wrong with a high fractional error if the value being computed itself flirts with the bounds of machine precision. In turn, these small absolute values of the integral typically correspond to regions where either $a$ or $b$ (or both) are very small. Since for many applications the integrals would be binned in $a$ and $b$ (in 3-D) after computation, the spherical-shell weights of $a^2b^2$ would greatly suppress the importance of any region of the integral our algorithm gets wrong. So we do not believe this to be a significant limitation in practice.

The main source of disagreement between the naive method and the rotation method seemed to be the large dynamic range of the 1-D integrands combined with the need for rather precise cancellation of a number of divergent such integrals to a finite sum. When one uses the rotation method, one has several 1-D integrals that have rather high powers of $k$ in the denominator (generically, they can be at most $\sim k^{-\ell - \ell'}$). Thus even a small integration interval in $k$ is converted into a large dynamic range of the integrand. We considered using a higher-precision numerical type (through \textsc{python's} \texttt{mpmath} library). However there are not many integration methods available in \textsc{python} for this type, and type conversion to it would increase runtimes as well. Consequently we deferred detailed investigation of its use. For the test cases we explored, the precision achieved already seemed sufficient for practical use.

\begin{figure}
    \centering
    \includegraphics[width=1.1\linewidth]{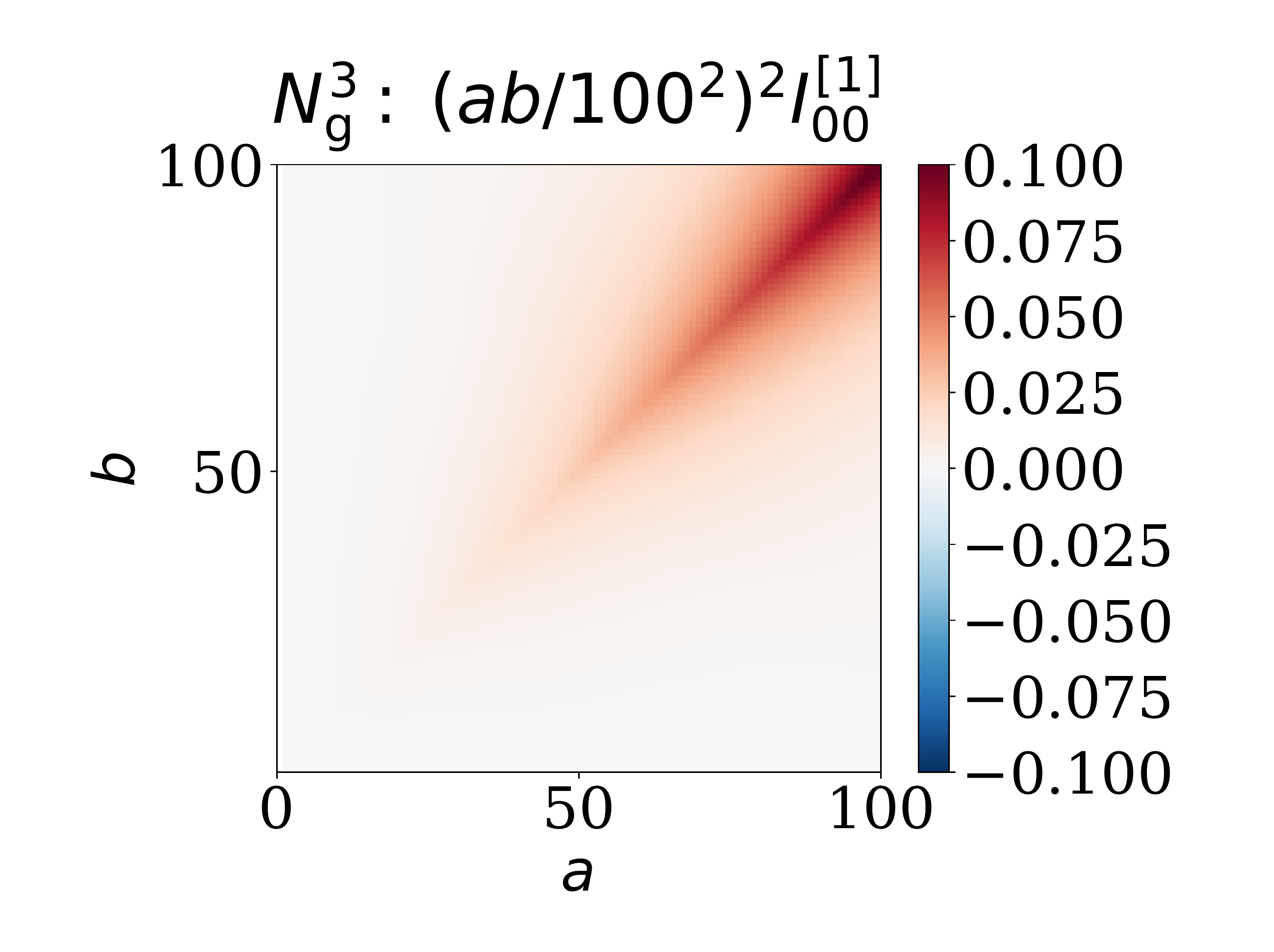}
    \caption{Here we show the weighted integral of $j_0(ka)j_0(kb)$ against $P(k)$ evaluated with the naive, $\Ng^3$ method. The weights correspond to those one would use when binning in $a$ and $b$ on spherical shells; the volume of a shell in $a,b$ scales as the bin width squared times $a^2b^2$. Our purpose here is simply to show the actual values to which subsequent plots of differences between the different methods should be compared.}
    \label{fig:j0j0_val}
\end{figure}

\begin{figure}
    \centering
    \includegraphics[width=1.1\linewidth]{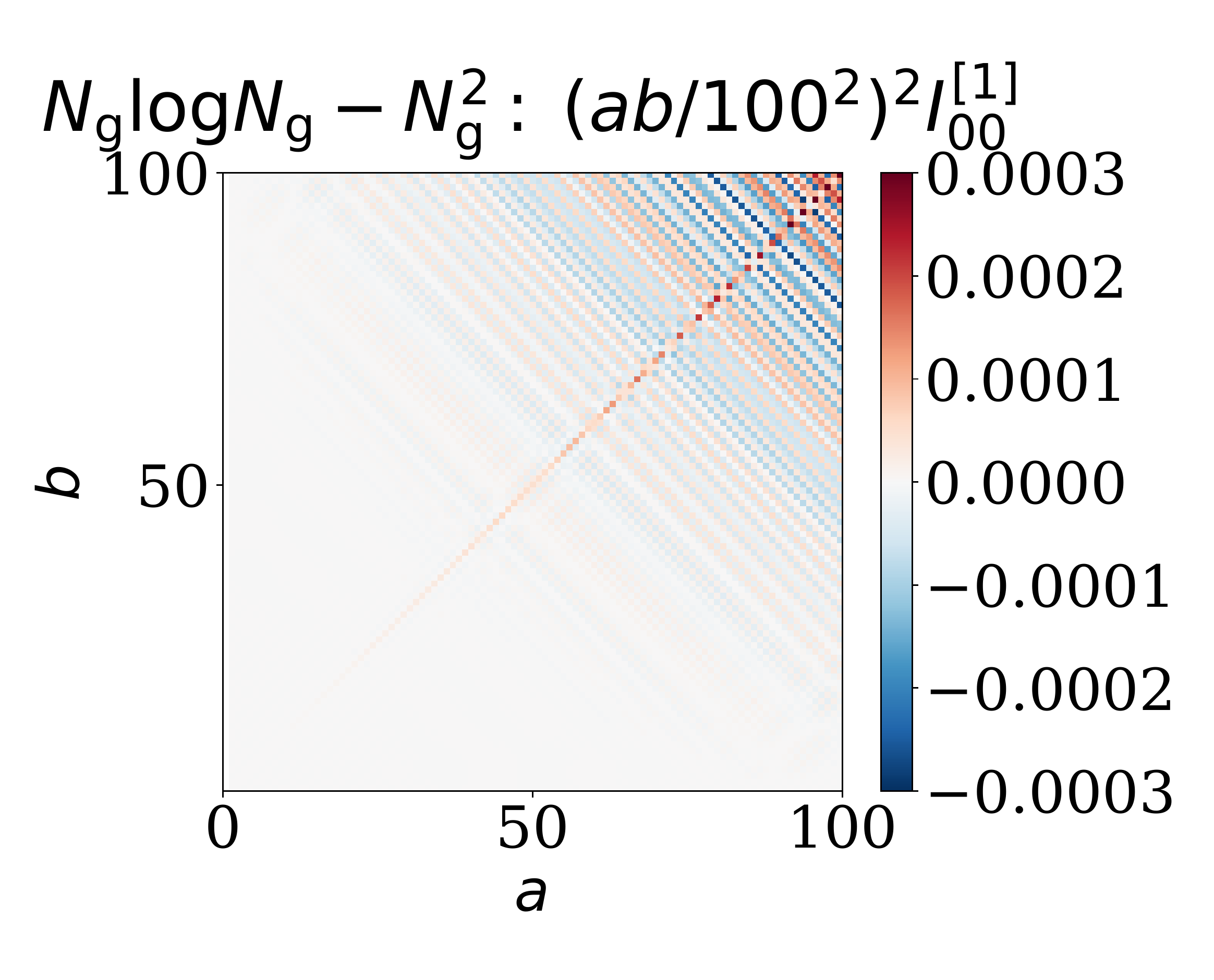}
    \caption{Here we show the weighted difference between the $\Ng \log \Ng$ and $\Ng^2$  methods. This plot should be compared to Figure \ref{fig:j0j0_val}; one can see that the differences between the two methods here are of order $1/300$ the integral's value. We note that $\Ng^2$ and $\Ng^3$ agree far better than this, though we do not show that plot for brevity. But as a consequence one can take it that this Figure also well-represents the difference between $\Ng \log \Ng$ and $\Ng^3$, as $\Ng^2$ can be taken as a proxy for the latter.}
    \label{fig:j0j0_diff}
\end{figure}

\begin{figure}
    \centering
    \includegraphics[width=1.1\linewidth]{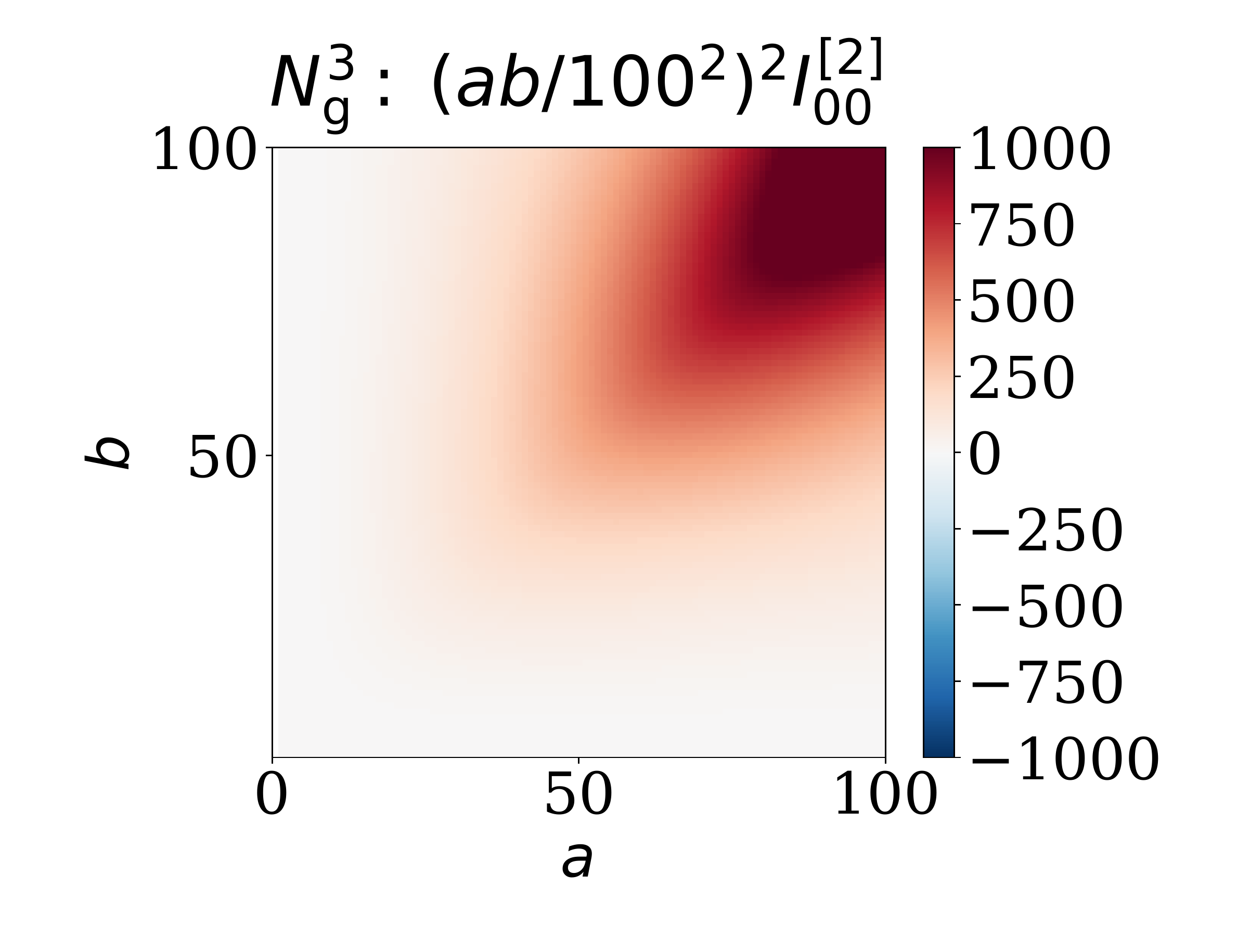}
    \caption{Here we show the weighted integral of $j_0(ka)j_0(kb)$ against $P^2(k)$ evaluated with the naive, $\Ng^3$ method. Our purpose here is simply to show the actual values to which subsequent plots of differences between the different methods should be compared. One can also compare this Figure to Figure \ref{fig:j0j0_val} to get a sense for how squaring the power spectrum alters the integral; essentially it extends the region of strong support of the result away from the diagonal. This makes sense as squaring the power spectrum is convolving the correlation function with itself. This is a smoothing operation and hence will make the off-diagonal contributions higher.}
    \label{fig:j0j0_Psq_val}
\end{figure}

\begin{figure}
    \centering
    \includegraphics[width=\linewidth]{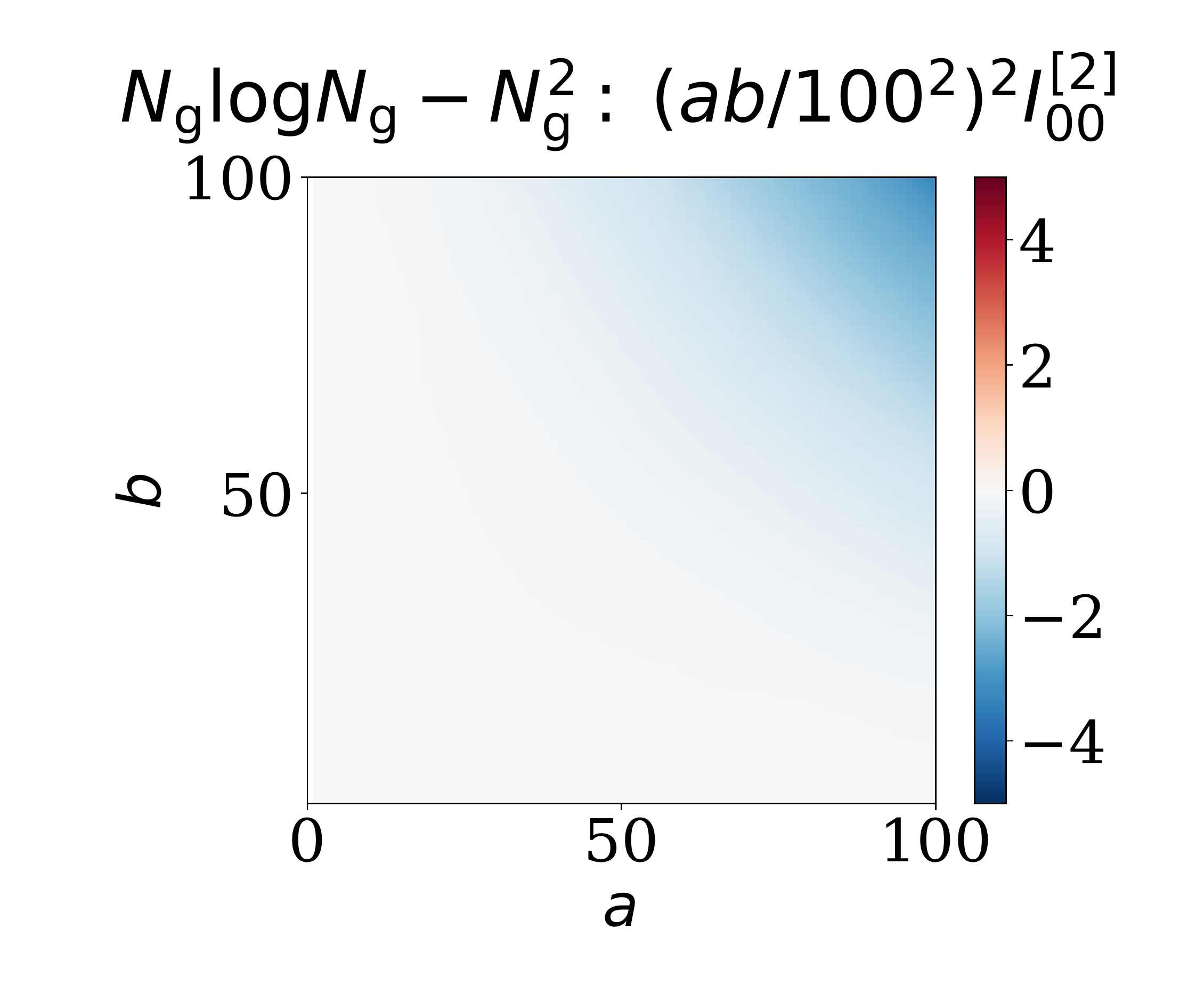}
    \caption{Here we show the weighted difference between the $\Ng \log \Ng$ and $\Ng^2$  methods for the integral of $j_0(ka) j_0(kb)$ against  $P^2$. This plot should be compared to Figure \ref{fig:j0j0_Psq_val}; one can see that the differences between the two methods here are of order $1/250$ the integral's value. We note that $\Ng^2$ and $\Ng^3$ agree far better than this, though we do not show that plot for brevity. But as a consequence one can take it that this Figure also well-represents the difference between $\Ng \log \Ng$ and $\Ng^3$, as $\Ng^2$ can be taken as a proxy for the latter.}
    \label{fig:j0j0_Psq_diff}
\end{figure}

\begin{figure}
    \centering
    \includegraphics[width=\linewidth]{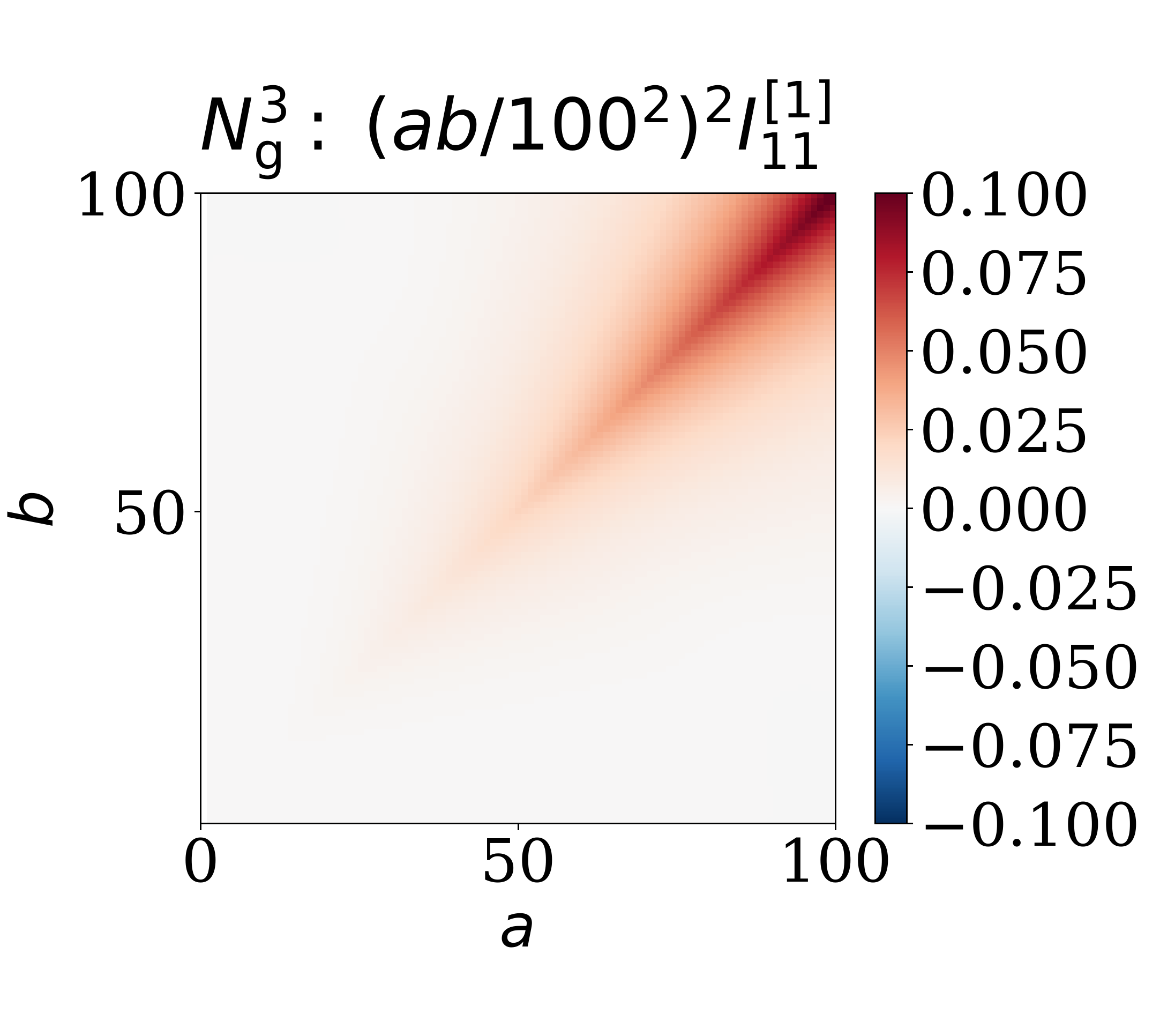}
    \caption{Here we show the weighted integral of $j_1(ka)j_1(kb)$ against $P(k)$ evaluated with the naive, $\Ng^3$ method. Our purpose here is simply to show the actual values to which subsequent plots of differences between the different methods should be compared. One can also compare this Figure to Figure \ref{fig:j0j0_val} to get a sense for how raising the order of each spherical Bessel function alters the integral; essentially it does not do so greatly. The structure remains primarily on the diagonal. }
    \label{fig:j1j1_val}
\end{figure}

\begin{figure}
    \centering
    \includegraphics[width=1.2\linewidth]{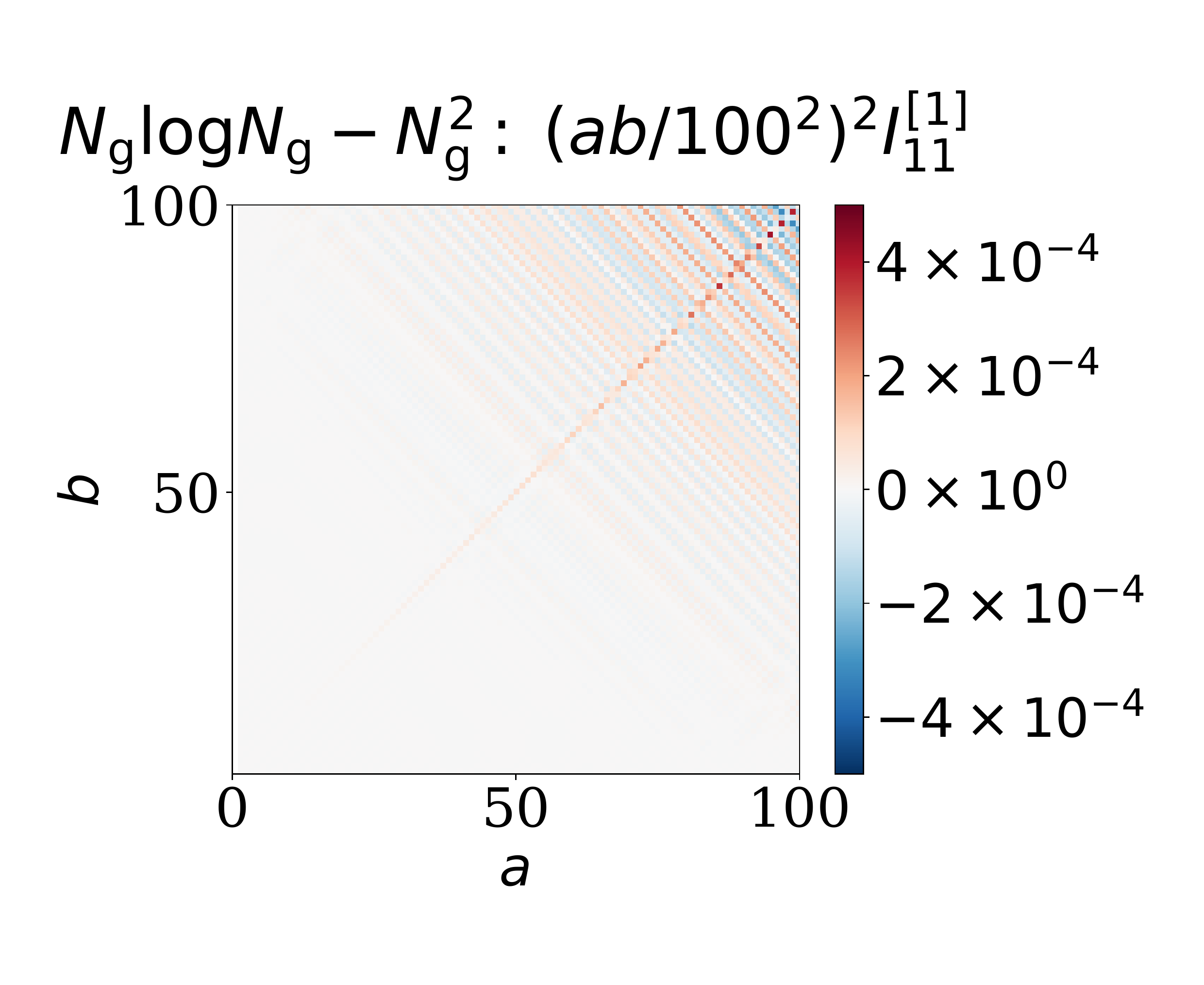}
    \caption{Here we show the weighted difference between the $\Ng \log \Ng$ and $\Ng^2$  methods for the integral of $j_1(ka) j_1(kb)$ against  $P$. This plot should be compared to Figure \ref{fig:j1j1_val}; one can see that the differences between the two methods here are of order $1/250$ the integral's value. We note that $\Ng^2$ and $\Ng^3$ agree far better than this, though we do not show that plot for brevity. But as a consequence one can take it that this Figure also well-represents the difference between $\Ng \log \Ng$ and $\Ng^3$, as $\Ng^2$ can be taken as a proxy for the latter.}
    \label{fig:j1j1_diff}
\end{figure}

\begin{figure}
    \centering
    \includegraphics[width=1.03\linewidth]{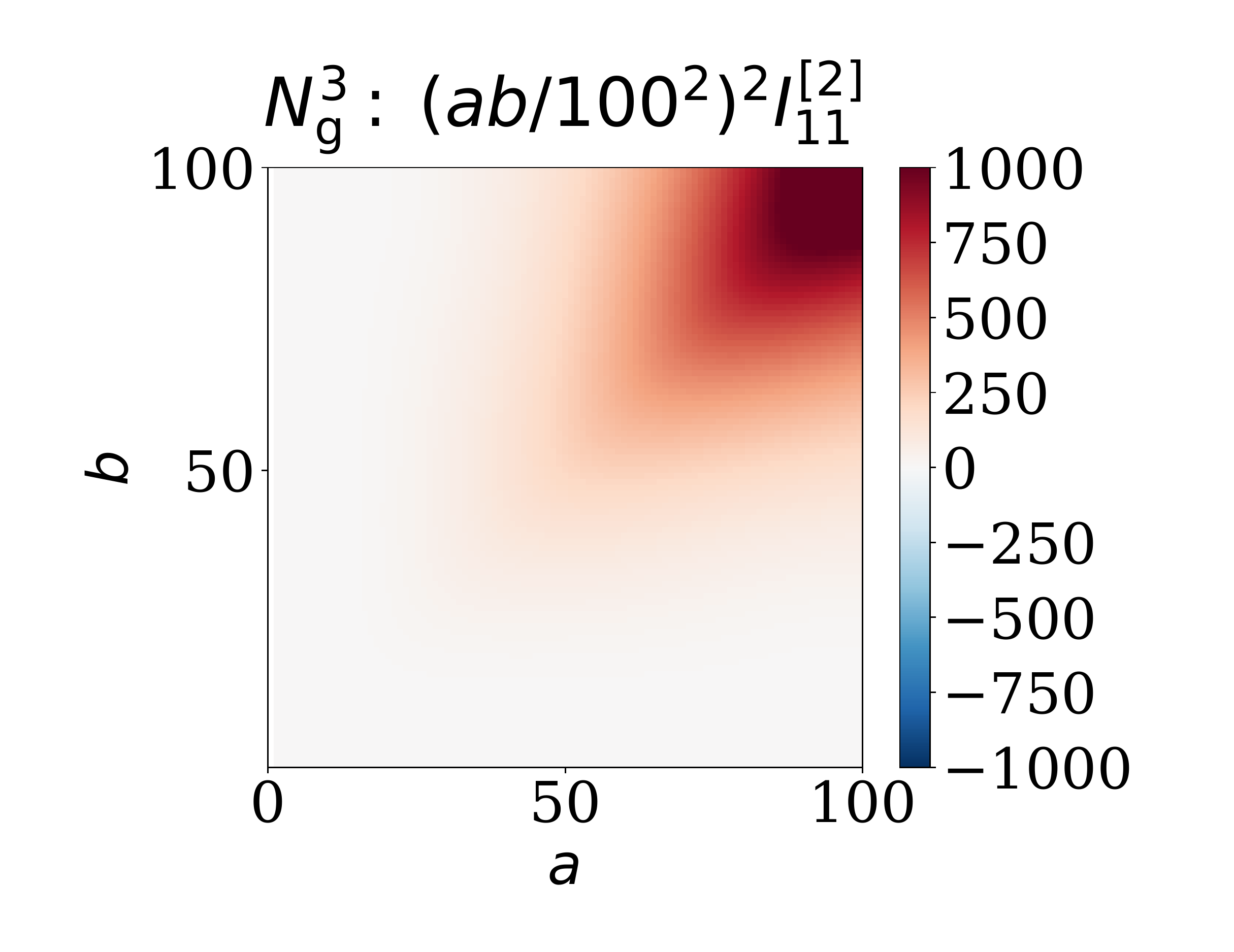}
    \caption{Here we show the weighted integral of $j_1(ka)j_1(kb)$ against $P^2(k)$ evaluated with the naive, $\Ng^3$ method. Our purpose here is simply to show the actual values to which subsequent plots of differences between the different methods should be compared. One can also compare this Figure to Figure \ref{fig:j1j1_val} to get a sense for how squaring the power spectrum alters the result; the take-away is the same as that discussed in the analogous case for $j_0(ka) j_0(kb)$ (Figure \ref{fig:j0j0_Psq_val}).}
    \label{fig:j1j1_Psq_val}
\end{figure}

\begin{figure}
    \centering
    \includegraphics[width=.95\linewidth]{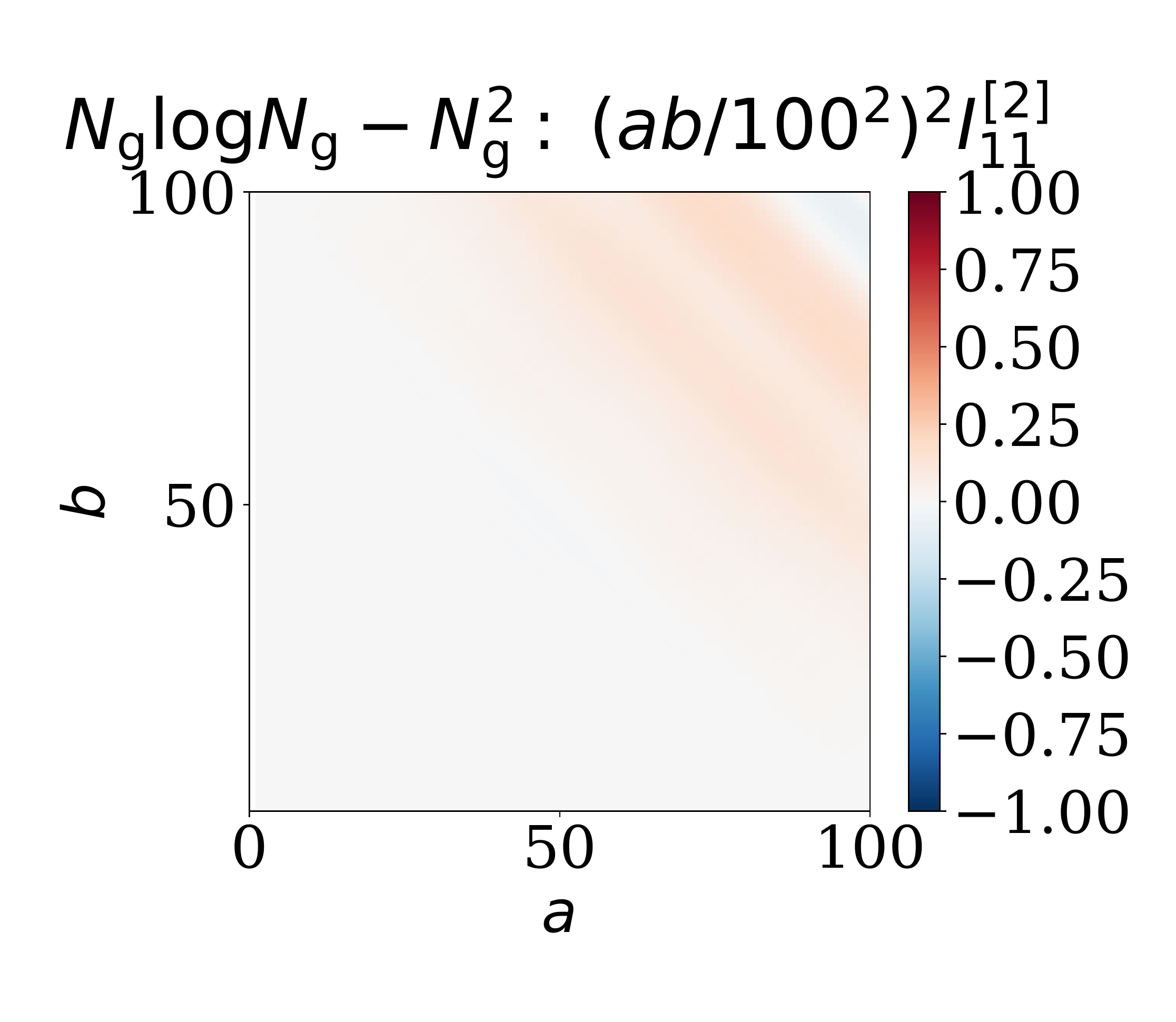}
    \caption{Here we show the weighted difference between the $\Ng \log \Ng$ and $\Ng^2$  methods for the integral of $j_1(ka) j_1(kb)$ against  $P^2(k)$. This plot should be compared to Figure \ref{fig:j1j1_Psq_val}; one can see that the differences between the two methods here are of order $1/1000$ the integral's value. We note that $\Ng^2$ and $\Ng^3$ agree far better than this, though we do not show that plot for brevity. But as a consequence one can take it that this Figure also well-represents the difference between $\Ng \log \Ng$ and $\Ng^3$, as $\Ng^2$ can be taken as a proxy for the latter.}
    \label{fig:j1j1_Psq_diff}
\end{figure}

\begin{figure}
    \centering
    \includegraphics[width=\linewidth]{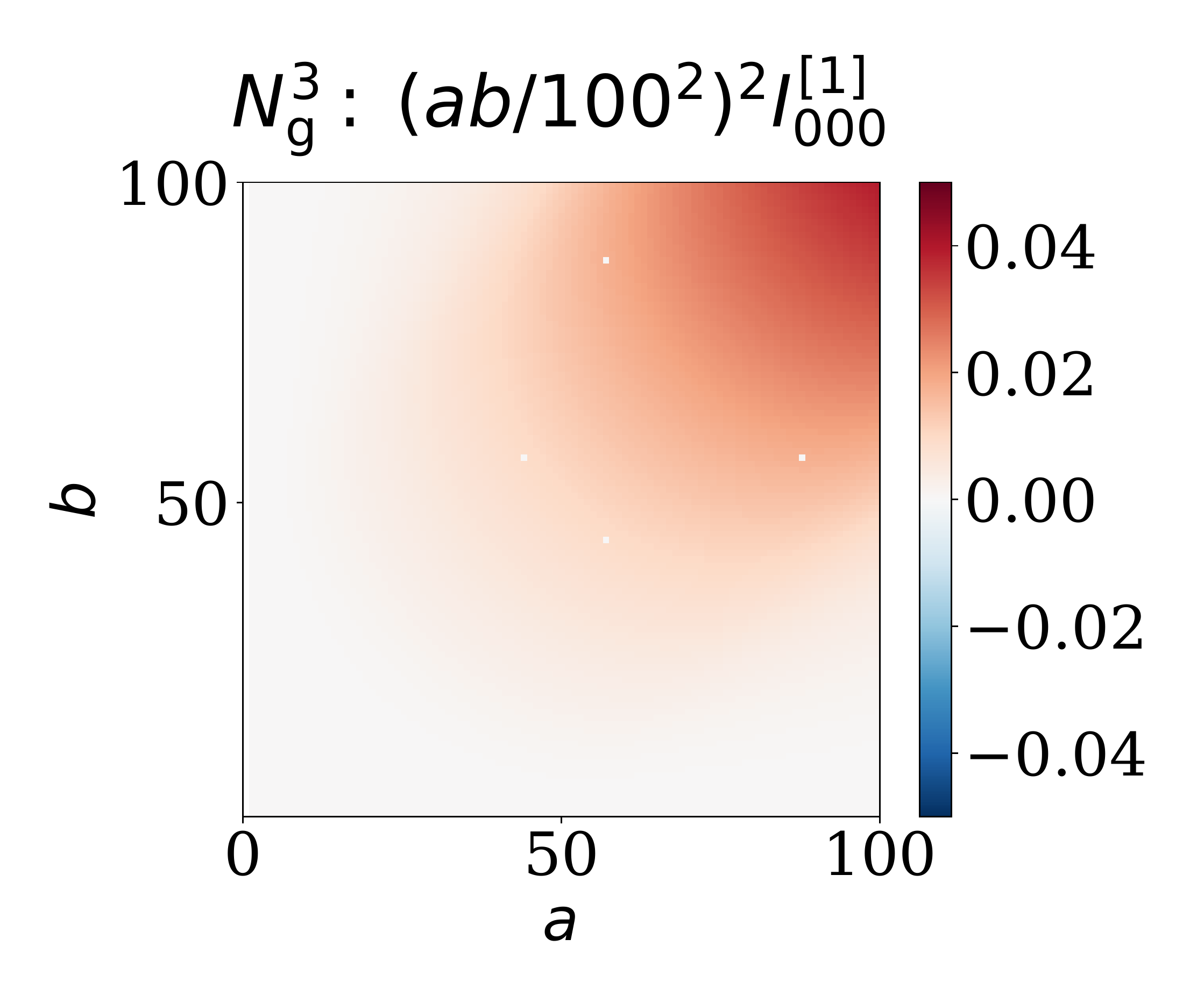}
    \caption{Here we show the weighted integral of $j_0(ka)j_0(kb) j_0(kc)$ against $P(k)$ evaluated with the naive, $\Ng^3$ method and displayed on a slice of constant $c = 50$. Our purpose here is simply to show the actual values to which subsequent plots of differences between the different methods should be compared. One can also observe that the structure is much more off diagonal than the results for the two-sBF integrals of both $P$ and $P^2$.}
    \label{fig:j0j0j0_val}
\end{figure}

\begin{figure}
    \centering
    \includegraphics[width=1.10\linewidth]{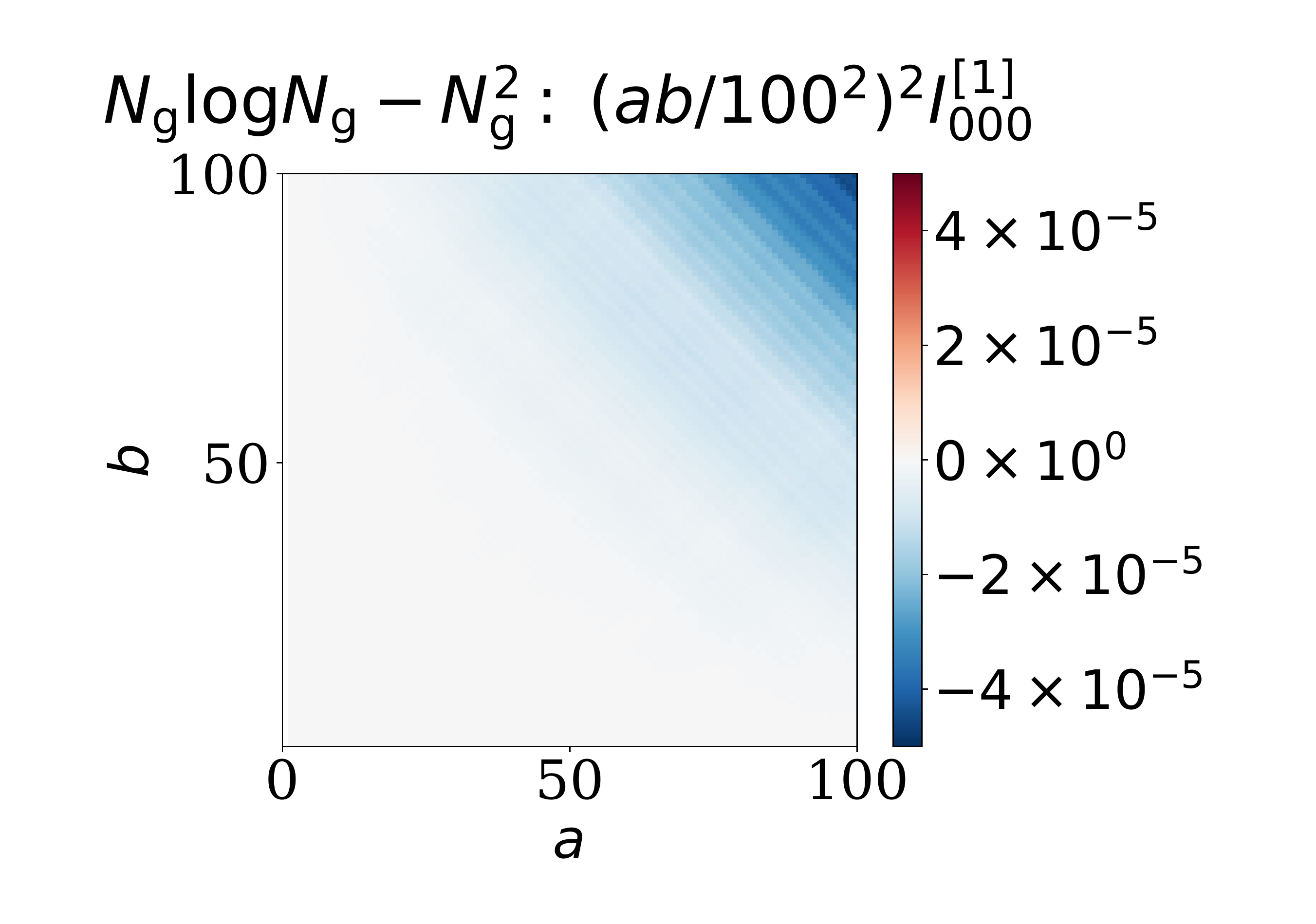}
    \caption{Here we show the weighted difference between the $\Ng \log \Ng$ and $\Ng^2$  methods for the integral of $j_0(ka) j_0(kb)$ against  $P(k)$ and displayed on a slice of constant $c=50$. This plot should be compared to Figure \ref{fig:j0j0j0_val}; one can see that the differences between the two methods here are of order $1/1000$ the integral's value. We note that $\Ng^2$ and $\Ng^3$ agree far better than this, though we do not show that plot for brevity. But as a consequence one can take it that this Figure also well-represents the difference between $\Ng \log \Ng$ and $\Ng^3$, as $\Ng^2$ can be taken as a proxy for the latter.}
    \label{fig:j0j0j0_diff}
\end{figure}

The overall comparisons for the precision and accuracy of the different methods can be read off from our Figures, which show the results with the naive method and then the differences between the $\Ng \log \Ng$ and $\Ng^2$ methods; since the $\Ng^3$ and $\Ng^2$ methods typically agreed much better than $\Ng \log \Ng$ and $\Ng^2$, the $\Ng \log \Ng$ differences from $\Ng^2$ can also be taken to indicate those between $\Ng \log \Ng$ and $\Ng^3$. Figures \ref{fig:j0j0_val} and \ref{fig:j0j0_diff} show the value and these differences for $I_{00}^{[1]}$; Figures \ref{fig:j0j0_Psq_val} and \ref{fig:j0j0_Psq_diff} show the value and these differences for $I_{00}^{[2]}$. Figures \ref{fig:j1j1_val} and \ref{fig:j1j1_diff} show the value and these differences for $I_{11}^{[1]}$; Figures \ref{fig:j1j1_Psq_val} and \ref{fig:j1j1_Psq_diff} show the value and these differences for $I_{11}^{[2]}$. Finally, Figures \ref{fig:j0j0j0_val} and \ref{fig:j0j0j0_diff} show the value and differences for $I_{000}^{[1]}$. We do not show plots for the $j_2j_2$ integrals because our results for them are extremely similar to those for the $j_0$ and $j_1$ cases.

\subsubsection{Timings}
\label{subsubsec:timings}
Regarding timings, we performed repeated tests of the $\Ng^2$ rotation method implementation with the Romberg method and on the fixed grid. We saw, averaged over 10 or so iterations, an acceleration of order $10-30\times$ over the naive, $\Ng^3$ method, with variation of about $10\%$. This acceleration factor should not be over-interpreted: the scaling of each method with the number of integration points is fundamentally different, so for different problems (e.g. a different number of $a$ and $b$ being integrated at, a different number of $k$ being integrated over), one would likely see different acceleration multipliers. This comment also holds true when considering the acceleration we found for the $\Ng \log \Ng$ over the $\Ng^2$ and $\Ng^3$. We typically found accelerations of $100\times$ for the $\Ng \log \Ng$ over the $\Ng^2$, and more like $1000-3000\times$ over the $\Ng^3$. We detail the absolute timings in the following subsections, but emphasize that the absolute timings will be hardware-dependent. Again we caution that these factors are problem-specific. However, the problems we evaluated are realistic use cases, so these factors likely do offer an accurate sense for the impact of our algorithm on e.g. DESI analytic covariance evaluations. We note that the more challenging the problem (i.e. the more $(a,b)$ or $k$ points needed), the more our FFTLog-based rotation-method implementation ($\Ng \log \Ng$) will outperform the others: its scaling with the number of input points is simply fundamentally better. This point should be borne in mind when choosing which algorithm to build into code that may then be scaled up.

\subsection{Unsmoothed power spectrum}
\label{subsec:unsmoothed}
We now go into greater detail on the numerical test cases we investigated. The primary use case we focused on testing in this work was the double-Bessel integrals relevant for the Gaussian Random Field contribution to the covariance matrix of the anisotropic 2PCF multipoles (e.g. \citealt{Xu:2013}), as might be used in a DESI analysis. Taking $P \to P + 1/n$ and then expanding the integral in equation (\ref{eqn:2pcf_covar}), we find three terms: one proportional to $P^2$, a second proportional to $P\times(1/n)$, and a third proportional to $(1/n)^2$, with $n$ the number density of the galaxy survey. The third may be done analytically, since $1/n$ is $k$-independent. The second and first must be done numerically, but of course the factor of $1/n$ in the second may also be pulled out.

So in short, for the covariance matrix, we will have numerical double-sBF integrals against $P$ and $P^2$. In a typical use-case, several additional modifications would be made relative to the test cases we examine.\\
\\
i) Smoothing of BAO in the power spectrum.\\
ii) Smoothing of the Fourier transform as we go to configuration space to avoid ringing.\\
iii) Binning over separations.\\
\\
We have already discussed these in more detail in \S\ref{sec:Covar}. We simply recapitulate them now to highlight that our test case, an ``unsmoothed'' power spectrum with i) no suppression of BAO ii) no smoothing of the FT, and iii) no binning, is a more stringent numerical test than the use cases that would likely occur. So the agreement we achieve here between the three methods tested should be taken as a conservative lower bound on the true agreement likely to occur in applications.

For our tests, we used a numerical power spectrum from CAMB (\citealt{Lewis:2000}) with a cosmology roughly that of Planck 2016 \citep{Planck:2016}. We also used this for the $P^2$ tests, described in \S\ref{subsec:squared}. For all of our test cases, we achieved very good agreement between the different implementations. For the $j_0 j_0$ test, the typical agreement was at the level of $10^{-16}$ between the naive and $\Ng^2$-rotation method, and the values were of order $10^{-1}$. The agreement between $\Ng^2$ and $\Ng \log \Ng$ rotation methods was at the level of $3\times 10^{-4}$. For the $j_1j_1$ test, the differences between naive and $\Ng^2$ were at the level of $10^{-9}$, with the values of the integrals of order $10^{-1}$ to $1$; differences between $\Ng^2$ and $\Ng \log \Ng$ were of order $4\times 10^{-4}$. The agreement between the three implementations was less good for the $j_2 j_2$ and $j_0 j_0 j_0$ tests but still typically better than $10^{-4}$ fractionally of the integrals' values.

For $I_{00}^{[1]}$, going from naive to $\Ng^2$ saw a factor of $30\times$ speedup ($9,200$ s vs. $360$ s for both integrals done with adaptive Romberg and a maximum of 15 refinements, 310 s vs. 10 s for both integrals done with adaptive Romberg and a maximum of 10 refinements). For $I_{11}^{[1]}$, the speedup was $10\times$ (30 s vs. 3 s for fixed grid, 370 vs. 32 s for adaptive). For $I_{22}^{[1]}$, the speedup was $7\times$ (390 s vs. 60 s, with adaptive Romberg). Going from $\Ng^2$ to $\Ng \log \Ng$, runtimes dropped to typically $0.2$ s; this corresponds to a factor of $100\times$ speed-up relative to the $\Ng^2$ and $1000-3000\times$ relative to the naive. The tests of the naive and $\Ng^2$ methods were run on a 2013 MacBook Air with Intel i7 processor and 8 GB 1600 MHz DDR3 RAM. The tests of the $\Ng \log \Ng$ method were run on a Lenovo Yoga 2 Pro with the same processor and amount of RAM, so the results can reasonably be cross-compared.

\subsection{Squared unsmoothed power spectrum}
\label{subsec:squared}
Squaring the power spectrum will more sharply concentrate it about its peak. This can also be intuitively seen since in configuration space, squaring would correspond to convolving the linear correlation function $\xi_0$ with itself. On the very simplistic cartoon model that $\xi_0$ is a Gaussian, convolving it with itself would increase the width by $\sqrt{2}$. By the uncertainty relation, we see that this would then shrink the width of the power spectrum by the same factor. This means that the range of $k$ over which $P^2$ is appreciable is smaller than that for $P$. Reducing this range should make our integration approach even more accurate. Indeed, the main source of inaccuracy is the high powers of $k$ in the denominators, where one needs to track them accurately to get good cancellation when they are summed up at the end of the rotation method. Reducing the range in $k$ over which the integrand is non-negligible will tend to make this issue better.

We tested for agreement between the three methods for $P^2$. We found sub-percent agreement as long as we used $k_{\rm max} = 4\;h/{\rm Mpc}$ for the $\Ng^3$ and $\Ng^2$ methods (the $k$ grid for the $\Ng \log \Ng$ method was logarithmically-spaced from $10^{-4}\;h/{\rm Mpc}$ to $100\;h/{\rm Mpc}$, with 8192 points. We did not extrapolate as a power-law outside either end of this range, which is also possible in this approach.

We believe the need to truncate at a different $k_{\rm max}$ to find good agreement between the methods is driven by details of how the $\Ng^3$ and $\Ng^2$ methods handle the integration. Both are done with adaptive Romberg, and the Romberg probably breaks down in accuracy when attempting to deal with the high-$k$ points where the sBFs (or sines and cosines) greatly oscillate. This likely drives it to spend most of its sampling points there, even though these high-$k$ regions do not matter for the value of the integrals. With adaptive Romberg, there is no way to control where it samples, save for the crude method of simply truncating at lower $k_{\rm max}$. With Romberg on a fixed grid, the grid must be linearly spaced, not log spaced, so using that approach is not an obvious fix. One could imagine remapping the integrand to be in terms of $u = \ln k$, and then the integrand would be linearly spaced in $u$ for a logarithmic $k$ grid. However, we did not pursue this, as we felt sub-percent agreement between the methods was sufficient. In any case, we advocate the $\Ng \log \Ng$ method for production analyses, and it is not affected by these points.

We note that the DESI statistical error bars on the signal should be of order $1\%$ in of order ten redshift bins \citep{DESI:2016}, for a total error of $0.1\%$ on the BAO scale when these are combined. We emphasize that these percentages are relative to the signal. Making a fractional error of $10^{-4}$ in the covariance (which by Taylor series translates to $\sim$$1/2\times 10^{-4}$ in the error bar itself) would then constitute a fraction $1/2\times 10^{-4}\times 1\%$ of the DESI signal. This level of error will be irrelevant to the derived likelihood.

\subsection{Analytic test case}
\begin{figure}
    \centering
    \includegraphics[width=\linewidth]{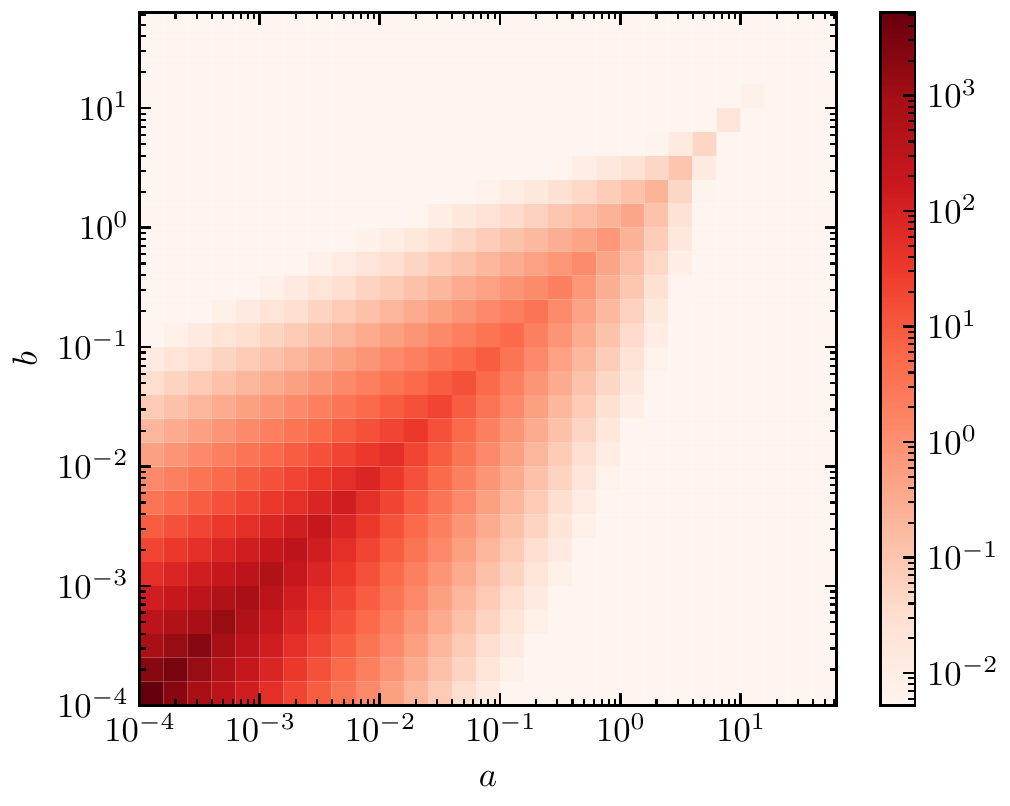}
    \caption{Here we show the integral for our analytic test case equation \eqref{test}, for comparison with the plots of the error in our method we then display in Figure \ref{fig:test_err}. This plot is just meant to give a sense for the structure of this integral to then aid in interpreting Figure \ref{fig:test_err}.}
    \label{fig:test}
\end{figure}

\begin{figure}
    \centering
    \includegraphics[width=\linewidth]{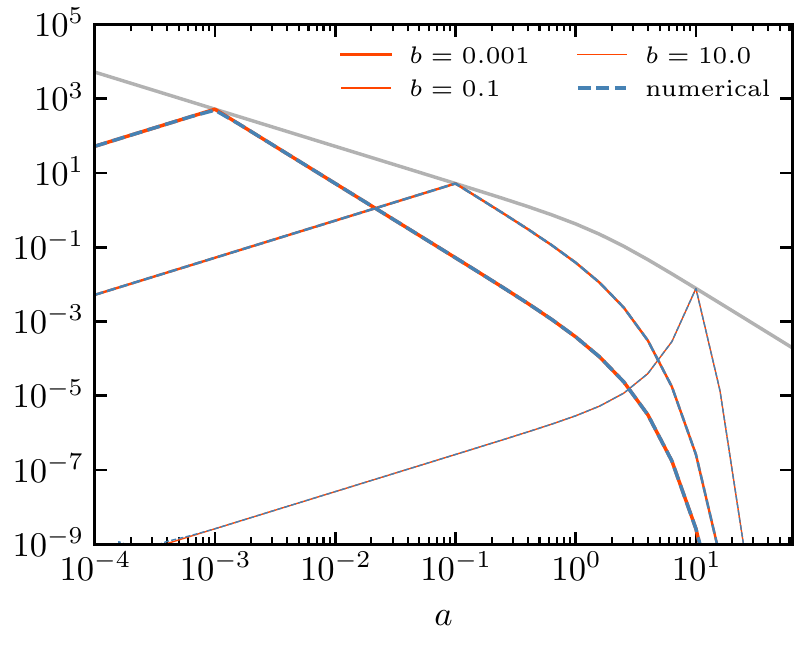}
    \caption{Comparison between numerical (dashed blue) and analytic (solid
    red) results, shown as functions of $a$ at the three different fixed values of $b$
   indicated in the legend and displayed in lines of varying width. The grey line gives the
    peak amplitudes on the trace $a=b$ through the $(a,b)$ plane, i.e. the diagonal of Fig.~\ref{fig:test}, for comparison. Of course the blue and red curves meet the grey when $a=b$; generally the amplitudes drop sharply away from the diagonal. Overall the agreement is quite good over a very large dynamic range of both $a$, $b$ and values of the integral.}
    \label{fig:test_slices}
\end{figure}

\begin{figure}
    \centering
    \includegraphics[width=\linewidth]{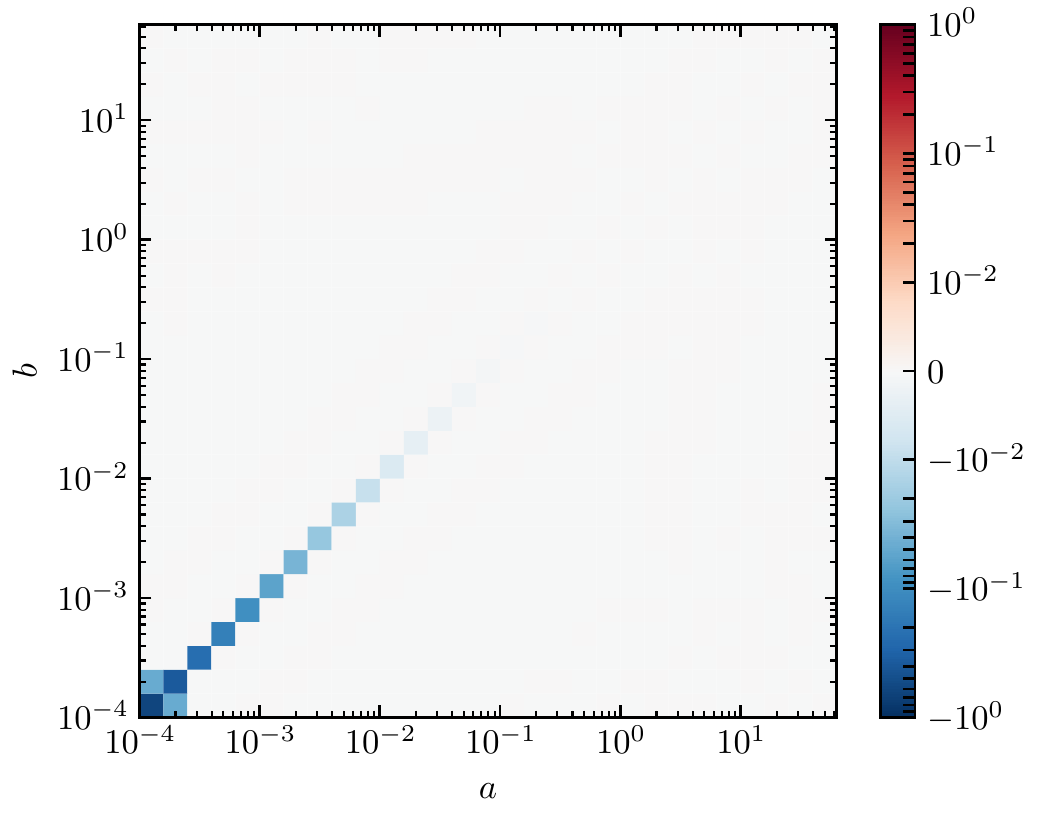}
    \caption{Numerical error, normalized by the square root of the product of the
    diagonal of Figure \ref{fig:test}, $\sqrt{I_{aa} I_{bb}}$. The residual at the lower left corner is at the 10\% level, consistent with
    Fig.~\ref{fig:test_slices} given the large range of the vertical axis there.
    }
    \label{fig:test_err}
\end{figure}

For additional confirmation of our method and implementation, we also evaluated a test case where the exact answer is known analytically. This test we focused on the FFTLog implementation of the rotation method, as that is the one we advocate putting into practical use as it is the fastest and has the best handling of dynamic range. Figures \ref{fig:test}-\ref{fig:test_err} show that the method performs quite well over a large range of $a$ and $b$. While it initially appears that the relative error becomes large in some regions, this is simply because machine precision sets a floor on the absolute error, while the absolute value of the integral becomes quite small. Thus the relative error goes up.

In practice, though, this rise in relative error will not matter. As we have already discussed, typically one bins the covariance (e.g. \citealt{Xu:2013}, equations 6-9), and the points where the absolute value of the integral is tiny will contribute very little to the binned result. Furthermore, many of the areas where the relative error is noticeable correspond to small $a$ or $b$, and when binning one weights by the bin volume, scaling as $a^2 b^2$. This weight will further reduce the  contribution these regions of higher relative error make to the binned covariance.

Finally, we note that the analytic test case is likely more stringent than the power spectrum (or its square) that might be typical use cases. The analytic test case is less sharply peaked than the power spectrum. The power spectrum scales roughly as $k/[1  + (k/k_{\rm eq})^2]^2$ and is usually smoothed by a Gaussian as $\exp[-k^2\sigma^2]$ with $\sigma \simeq 1$ Mpc.\footnote{$k_{\rm eq}$ is the wavenumber corresponding to the scale of the horizon at matter-radiation equality, and this term gives the turnover of the matter transfer function for modes that entered the horizon during radiation-domination and thus had their growth suppressed as the potential evolved. This was first noted in \cite{Meszaros:1974}; a configuration-space picture is presented in \cite{Slepian_TF:2016}. Also in the numerator we took it that the scalar spectral tilt $n_s \simeq 1$; in reality it is measured to be $n_s = 0.965 \pm 0.004$ from $Planck$ data \citep{Planck:2018}.} The power spectrum scaling is thus, quite roughly, the square of our analytic test case, making the region where the power spectrum is non-negligible rather narrower than that for the analytic test case. Thus for the actual power spectrum, we need to resolve less dynamic range in $k$ than we do for the analytic test case, making this latter a more stringent test of our method.

We now detail our analytic test case. It is \cite{Gradshteyn:2007} 6.577.1,
\begin{align}
\label{577}
&\int_0^\infty x^{\nu-\mu+1+2n} J_\mu(ax) J_\nu(bx) \frac{d x}{x^2+c^2}\\
&  = (-1)^n c^{\nu-\mu+2n} I_\mu(ac) K_\nu(bc),\;\;\;{\rm for}  \nonumber\\
&b>a>0, \Re c>0, 2+\Re\mu-2n > \Re\nu > -1-n, 0\leq n\in\mathbb{Z},\nonumber
\end{align}
with the following special case when $n=0$, $\mu=3/2$, and $\nu=3/2$:
\begin{multline}
    \label{test}
    \int_0^\infty j_1(ax) j_1(bx) \frac{x^2\,\d x}{1+x^2}
    = \frac\pi{2a^2b^2}   \\\
    \times
    \begin{cases}
        (1+b) e^{-b} (a\cosh a - \sinh a), \quad a\leq b,  \\
        (1+a) e^{-a} (b\cosh b - \sinh b), \quad a>b.
    \end{cases}
\end{multline}

Expanding the lefthand side of equation \eqref{test}, we find eight terms (as displayed in equation \ref{eqn:j1j1_expansion}, and setting $\Sigma = a + b$ and $\Delta = a -b$). Two out of these terms are proportional to
\begin{equation*}
    \int_0^\infty\! d x\, \frac{\cos[(a\pm b) x]}{x^2(1+x^2)}
\end{equation*}
which diverges in the infrared $(k\to 0)$. However their sum converges. We will use \texttt{mcfit} to compute an incorrect finite result
for each of the divergent terms and expect they will sum up to the correct one (as the true, divergent results would also do).

\texttt{mcfit} is a \textsc{Python} package that implements the FFTLog algorithm for
fast Hankel transforms (\citealt{Talman:1978}, \citealt{Hamilton:2000}).
The algorithm assumes that the input function is sampled on a logarithmically
spaced grid.
It then approximates the input function (of logarithmic variable) with a
truncated Fourier series, and evaluates the exact Hankel transform of each
Fourier mode, before adding them up.
The analytic treatment avoids numerically computing the oscillatory integrals
involving Bessel functions, therefore leading to more accurate and efficient
results.
We refer the readers to the above references for more details.

We use \texttt{mcfit} to integrate each term of the expansion equation (\ref{eqn:j1j1_expansion}). Each term can be computed to give a finite value, even the seemingly divergent ones.
This is because the exact Hankel transform of the Fourier modes is in a form
whose divergence can be removed by analytic continuation.\footnote{The explicit
form is equation (B18) of \cite{Hamilton:2000}, where divergences
arise at the poles of the Gamma function.}

The rotation method requires a linear grid, which can be obtained by
interpolation from the logarithmically-spaced output of \texttt{mcfit}.
However, interpolations, typically with cubic spline,  are prone to numerical
errors, especially so when one output variable is much larger than the other.
We define $C_n$ and $S_n$ as follows:
\begin{align}
    C_n(y) &= \int \frac{x^2\,d x}{1+x^2} \frac{\cos(xy)}{x^n}, \\
    S_n(y) &= \int \frac{x^2\, d x}{1+x^2} \frac{\sin(xy)}{x^n}.
\end{align}
If $a \gg b$, combinations like $C_n(a+b) - C_n(a-b)$ would produce $2b \, C_n'(a)$ at leading order, which involves the derivative of the spline. As a numerical derivative, this will generally be less accurate than the spline interpolation itself. To solve this problem, we adopt a Hermite spline in which the derivatives are
explicitly supplied alongside the function values.
With a Hermite spline, the piecewise polynomials are determined locally by the
function's values, its derivatives, and the end points.
This is in contrast with the (cubic) interpolating spline, whose piecewise
polynomials are determined globally by all the function values.
Specifially for the test case we use the quintic Hermite spline that contructs
piecewise polynomials with both first and second derivatives.
The derivatives of $C_n$ and $S_n$ form a mutual recursion
\begin{align}
&\frac{d S_n}{d y} = C_{n-1}, \nonumber\\
&\frac{d C_n}{d y} = - S_{n-1}.
\end{align}
where on the righthand side the derivatives are terms that also appear in the
rotation method expansion. This allows us to reuse those numerical computations in contructing the Hermite
splines. We have verified that the quintic Hermite spline gives more accurate results
than the cubic interpolating spline.

To compute each $C_n$ or $S_n$, we sample $1/(1 + x^2)$ from $k = 10^{-2}\;h/{\rm Mpc}$ to
$10^4\;h/{\rm Mpc}$ at 300 logarithmically-spaced points, before extending them on both
sides using power-law extrapolations from the respective end segments. The extrapolation is implemented in \texttt{mcfit} as a default option, and the
total number of data points including the extrapolation is 8192.
\texttt{mcfit} then performs Fourier sine or cosine tranforms to compute $S_n$
or $C_n$, before adding them up to obtain the result.
On a single core Intel i7-4510U CPU ($2.00$ GHz), the numerical computation
with quintic Hermite interpolation takes about 200 ms.
However, this is obviously limited and dominated by the \texttt{scipy}
implementation of Hermite spline
(\texttt{scipy.interpolate.BPoly.from\_derivatives}), given that computation
with a cubic spline only takes about 30 ms.

\section{Conclusion \& Discussion}
\label{sec:concs}
In this work, we have presented a new way of evaluating integrals of two or more spherical Bessel functions against a numerical source function such as the matter power spectrum. This is a common problem in cosmology, as these integrals generically emerge when correlation functions with both points displaced from the origin are written as a multipole series. Our method works by rewriting the spherical Bessel functions as sines and cosines weighted by powerlaws, multiplying them out, and recoupling them into sines and cosines of the sum and difference of frequencies. This corresponds to rotating 45 degrees in the plane of free frequencies (i.e. not the integration variable), and essentially decouples the problem so that 3-D integrals become a sum of 2-D integrals. The method also trivially extends to integrals of three or more sBFs, as we show with the example of $j_0(ka)j_0(kb)j_0(kc)$. While the method costs more integrals, they are all 2-D, and this results in a signficant computational cost savings over the naive approach to this computation. In particular, if FFTLog is used to evaluate the required 2-D integrals, then they become even faster than 2-D. They then scale as $\Ng \log \Ng$ as compared to $\Ng^2$ for the ``rotation'' approach without FFTLog and $\Ng^3$ for the ``naive'' approach. In our implementation of a typical use case, the integral of two sBFs against the power spectrum, this yields a roughly $1000-3000\times$ speed-up (depending on the specific combination of sBF indices and other details). Unlike other work, our method does not require any expansion or approximation of the power spectrum (e.g. into complex power laws). The direct ($\Ng^2$) version of our method can also be used on any desired numerical grid, whereas the power law method must use a regular grid for its FT. Our method is also the first to show how to perform these integrals for mismatched indices (i.e. $\ell \neq \ell'$ for the sBFs) and for three or more sBFs.

The most closely related works to the present are \cite{Assassi:2017} and \cite{Gebhardt:2018}, also discussed in \S\ref{sec:accel}. These works expand the power spectrum in complex power laws (which can capture oscillatory behavior) and use analytic results for the integrals of 2 sBFs against a power law to perform the integrations. The key requirement of those works is that the power spectrum can be expanded as a sum of complex power laws; this is just the condition that it has an inverse Fourier Transform, as the expansion coefficients are in fact its discrete inverse FT. This should hold for most standard power spectra but will fail for ones that diverge in the IR faster than $k^{-2}$. Local non-Gaussianity generates a nearly-IR-divergent scale-dependent bias as $k^{-2}$ \citep{Dalal:2008}, so we speculate that more exotic non-Gaussianity models might diverge faster. They would therefore not interact well with the power-law method. However, they might still convergently be integrated against double sBFs with $\ell$ or $\ell'$ greater than zero because these will cancel the divergence. Our method could thus be used in this case even though the power-law method would break down.

Regarding computational efficiency, the power-law method requires one inverse FT to get the power spectrum's coefficients, and can then use a pre-computed lookup table of the analytic results for the power-law-double-sBF integrals and perform a matrix multiply with the coefficients. Creating the lookup table would be expensive but need only be done once. The lookup table would then need to be loaded from disk, which might be slow depending on the read rate and size of the table. In contrast, our method does not require any lookup tables, but computes a small number of FTs on the fly, all of which can remain in RAM during the computation. The speed comparison between the power law method and ours would thus be machine-dependent because each uses different pieces of the machine. 

The approach presented here should be highly useful for problems such as computing the analytic, Gaussian Random Field template for the anisotropic 2PCF multipoles' covariance matrix, such as might be required in analyzing large redshift-survey data. In particular, future surveys are likely to be analyzed using MCMC over cosmological parameters and one would like to recompute the covariance matrix at each set of parameters. Thus one might be computing the covariance matrix millions of times, making efficient evaluation of it highly desirable. We have tested all the numerical cases required for the covariance matrix of the monopole and quadrupole, as well as cross-terms. We find agreement between our new method and the naive approach sufficient so that any differences are a negligible source of systematic error in e.g. a DESI analysis.

The approach presented here needs more work to extend to higher sBF index $(\ell)$. This is because the recoupled, 2-D integrals it uses have large powers of $k$ in their denominators, which produce substantial dynamic range in the integrand and thereby strain the bounds of numerical precision. In particular, the powers of $k$ grow with the maximal $\ell$ used in the computation, and thus a small input range of $k$ maps to a larger and larger range for the integrand. Some of the integrals entering the expansion become formally divergent, though these divergences of course cancel when the integrals are summed in the end. The divergences do not appear numerically because we always integrate over a finite range. However in practice one still ends up needing quite precise cancellations between large values to get a smaller end result. We considered several schemes to address this issue, but none worked satisfactorily (and indeed, none were needed for the cases on which we focused). Future work might explore if splitting the broadband behavior of the power spectrum from the BAO wiggles helps alleviate this issue. The BAO wiggles are only present over roughly two decades in $k$, and one could imagine treating the broadband as a power law over much of its range, performing those integrals analytically, and using our scheme only for the BAO piece. Restricting the input range would of course reduce the output dynamic range in the integrand. However, even this approach would only go so far as $\ell$ rises, since even two decades produces an intractable dynamic range if raised to a sufficiently high power.

Nonetheless, overall this paper represents a new, independent, highly efficient method for evaluating the types of double and triple-sBF integrals ubiquitous in cosmology, and for many use-cases the $\ell$ we have already implemented are sufficient. We hope this algorithm will be of use in analyzing data from the next generation of large surveys.

\section*{Acknowledgements}
ZS thanks Donghui Jeong and Henry Grasshorn Gebhardt for sharing some test files. ZS also thanks Bob Cahn, Daniel J. Eisenstein, Emanuele Castorina, Marko Simonovi\'c, and Martin J. White for useful conversations. Support for some of this work was provided by the National Aeronautics and Space Administration through Einstein Postdoctoral Fellowship Award Number PF7-180167 issued by the Chandra X-ray Observatory Center, which is operated by the Smithsonian Astrophysical Observatory for and on behalf of the National Aeronautics Space Administration under contract NAS8-03060. ZS also acknowledges financial support from a Chamberlain Fellowship at Lawrence Berkeley National Laboratory (held prior to the Einstein) and from the Berkeley Center for Cosmological Physics. YL acknowledges support from Fellowships at the Berkeley Center for
Cosmological Physics, and at the Kavli IPMU established by World Premier
International Research Center Initiative (WPI) of the MEXT, Japan.
MS acknowledges support from the Corning Glass Works Fellowship at IAS as well as the National Science Foundation.




\bibliographystyle{mnras}
\bibliography{jfn_bib}



\appendix
\onecolumn
\section{Product of two spherical Bessel functions in terms of trigonometric functions}
\label{app:ExplicitJlJlExpansion}
From \cite{NIST_DLMF} 10.49.2 we have
\begin{align}
  \label{eq:1}
  j_\ell(kr) =
\sin\left(kr-\frac{\ell\pi}{2}\right)
\sum_{n=0}^{\myfloor{\ell/2}} (-1)^n
\frac{a_{2n}(\ell+\frac{1}{2})}{(kr)^{2n+1}}
+
\cos\left(kr-\frac{\ell\pi}{2}\right)
\sum_{m=0}^{\myfloor{(\ell-1)/2}} (-1)^m
\frac{a_{2m+1}(\ell+\frac{1}{2})}{(kr)^{2m+2}},
\end{align}
where
\begin{align}
  \label{eq:2}
  a_w(\ell+\tfrac{1}{2}) =
\frac{
[4(\ell+\frac{1}{2})^2-1^2]
[4(\ell+\frac{1}{2})^2-3^2]
\cdots
[4(\ell+\frac{1}{2})^2-(2w-1)^2]
}
{w!\;8^w}
\end{align}
is a polynomial in $\ell$.

The product of two spherical Bessel functions is then
\begin{align}
  j_\ell(kr)j_{\ell'}(kr') = &
\sin\left(kr-\frac{\ell\pi}{2}\right)
\sin\left(kr'-\frac{\ell'\pi}{2}\right)
\sum_{n=0}^{\myfloor{\ell/2}}
\sum_{n'=0}^{\myfloor{\ell'/2}}
(-1)^{n+n'}
\frac{a_{2n}(\ell+\frac{1}{2})}{(kr)^{2n+1}}
\frac{a_{2n'}(\ell'+\frac{1}{2})}{(kr')^{2n'+1}}
\nonumber\\
&+
\sin\left(kr-\frac{\ell\pi}{2}\right)
\cos\left(kr'-\frac{\ell'\pi}{2}\right)
\sum_{n=0}^{\myfloor{\ell/2}}
\sum_{m'=0}^{\myfloor{(\ell'-1)/2}}
(-1)^{n+m'}
\frac{a_{2n}(\ell+\frac{1}{2})}{(kr)^{2n+1}}
\frac{a_{2m'+1}(\ell'+\frac{1}{2})}{(kr')^{2m'+2}}
\nonumber\\
&+
\cos\left(kr-\frac{\ell\pi}{2}\right)
\sin\left(kr'-\frac{\ell'\pi}{2}\right)
\sum_{m=0}^{\myfloor{(\ell-1)/2}}
\sum_{n'=0}^{\myfloor{\ell'/2}}
(-1)^{m+n'}
\frac{a_{2m+1}(\ell+\frac{1}{2})}{(kr)^{2m+2}}
\frac{a_{2n'}(\ell'+\frac{1}{2})}{(kr')^{2n'+1}}
\nonumber\\
&+
\cos\left(kr-\frac{\ell\pi}{2}\right)
\cos\left(kr'-\frac{\ell'\pi}{2}\right)
\sum_{m=0}^{\myfloor{(\ell-1)/2}}
\sum_{m'=0}^{\myfloor{(\ell'-1)/2}}
(-1)^{m+m'}
\frac{a_{2m+1}(\ell+\frac{1}{2})}{(kr)^{2m+2}}
\frac{a_{2m'+1}(\ell'+\frac{1}{2})}{(kr')^{2m'+2}}.
\label{eqn:double_sbf_identity}
\end{align}
We can manipulate these relations to obtain explicit reduction formulae for an arbitrary product of any number of spherical Bessel functions. Here we obtain results for products of two; it is trivial to pursue the same approach for products of three if desired.

We first take each trigonometric function and use the angle sum formulae to rewrite it as a sum of products of trigonometric functions of $kr$, $\ell \pi/2$, $kr'$ and $\ell' \pi /2$. For instance, for the first term we find
\begin{align}
\sin\left(kr-\frac{\ell\pi}{2}\right)
\sin\left(kr'-\frac{\ell'\pi}{2}\right)=&\sin kr \sin kr' \cos \frac{\ell \pi}{2} \cos\frac{\ell' \pi}{2}
-\sin kr \cos kr' \cos \frac{\ell \pi}{2} \sin\frac{\ell' \pi}{2}
-\cos kr \sin kr' \sin \frac{\ell \pi}{2} \cos \frac{\ell' \pi}{2}\nonumber\\
&+\cos kr \cos kr' \sin \frac{\ell \pi}{2} \sin\frac{\ell' \pi}{2}.
\end{align}
We now notice that
\begin{align}
\cos \frac{\ell\pi}{2} = (1-\ell\; {\rm mod}\;2)(-1)^{\ell/2} \equiv K_{\ell}^{\rm c}.
\end{align}
We then have the product
\begin{align}
\cos \frac{\ell \pi}{2} \cos \frac{\ell' \pi}{2} \equiv K_{\ell \ell'}^{\rm c c}= K_{\ell}^{\rm c}  K_{\ell'}^{\rm c}.
\end{align}
This notation enables us to write
\begin{align}
\sin\left(kr-\frac{\ell\pi}{2}\right)
\sin\left(kr'-\frac{\ell'\pi}{2}\right) = \frac{1}{2}\bigg\{\left[\cos k\Delta -\cos k\Sigma\right]K_{\ell \ell'}^{\rm cc}
-\left[\sin k\Sigma + \sin k\Delta\right]K_{\ell \ell'}^{\rm cs}
-\left[\sin k\Sigma - \sin k\Delta\right]K_{\ell \ell'}^{\rm sc}
-\left[\cos k\Delta + \cos k\Sigma\right]K_{\ell \ell'}^{\rm ss}\bigg\}.
\end{align}
We may obtain similar results for the other terms in equation (\ref{eqn:double_sbf_identity}). We find
\begin{align}
\sin\left(kr-\frac{\ell\pi}{2}\right)
\cos\left(kr'-\frac{\ell'\pi}{2}\right) = \frac{1}{2}\bigg\{\left[\sin k\Sigma +\sin k\Delta\right]K_{\ell \ell'}^{\rm cc}
+\left[\cos k\Delta - \cos k\Sigma\right]K_{\ell \ell'}^{\rm cs}
-\left[\cos k\Delta + \cos k\Sigma\right]K_{\ell \ell'}^{\rm sc}
-\left[\sin k\Sigma - \sin k\Delta\right]K_{\ell \ell'}^{\rm ss}\bigg\}.
\label{eqn:A8}
\end{align}
The third term in equation (\ref{eqn:double_sbf_identity}) is just the second term with $r \leftrightarrow r'$ and $\ell \leftrightarrow \ell'$, so we have $\Sigma \to \Sigma$, $\Delta \to -\Delta$, $\ell \to \ell'$, $\ell' \to \ell$ in equation (\ref{eqn:A8}), yielding
\begin{align}
\cos\left(kr-\frac{\ell\pi}{2}\right)
\sin\left(kr'-\frac{\ell'\pi}{2}\right) = \frac{1}{2}\bigg\{\left[\sin k\Sigma - \sin k\Delta\right]K_{\ell \ell'}^{\rm cc}
+\left[\cos k\Delta - \cos k\Sigma\right]K_{\ell \ell'}^{\rm cs}
-\left[\cos k\Delta + \cos k\Sigma\right]K_{\ell \ell'}^{\rm sc}
-\left[\sin k\Sigma + \sin k\Delta\right]K_{\ell \ell'}^{\rm ss}\bigg\}.
\end{align}
Finally, for the last term in equation (\ref{eqn:double_sbf_identity}) we find
\begin{align}
\cos\left(kr-\frac{\ell\pi}{2}\right)
\cos\left(kr'-\frac{\ell'\pi}{2}\right) = \frac{1}{2}\bigg\{\left[\cos k\Delta + \cos k\Sigma\right]K_{\ell \ell'}^{\rm cc}
+\left[\sin k\Sigma - \sin k\Delta\right]K_{\ell \ell'}^{\rm cs}
+\left[\sin k\Sigma + \sin k\Delta\right]K_{\ell \ell'}^{\rm sc}
+\left[\cos k\Delta - \cos k\Sigma\right]K_{\ell \ell'}^{\rm ss}\bigg\}.
\end{align}
We now want to sum the four terms, each weighted by the double sum given in equation (\ref{eqn:double_sbf_identity}). Denoting the double sums in equation (\ref{eqn:double_sbf_identity}) by $w_1$ through $w_4$ we may co-add all terms in each trigonometric function of $k\Delta$ and $k\Sigma$, finding
\begin{align}
j_{\ell}(kr)j_{\ell'}(kr') =&\frac{1}{2}\bigg\{
\cos k\Delta \bigg[ w_1(K_{\ell \ell'}^{\rm cc} + K_{\ell \ell'}^{\rm ss})  +w_2(K_{\ell \ell'}^{\rm cs} - K_{\ell \ell'}^{\rm sc})
+w_3(K_{\ell \ell'}^{\rm cs} - K_{\ell \ell'}^{\rm sc})
+w_4(K_{\ell \ell'}^{\rm cc} + K_{\ell \ell'}^{\rm ss})\bigg]\nonumber\\
&+\cos k\Sigma \bigg[ w_1(-K_{\ell \ell'}^{\rm cc} + K_{\ell \ell'}^{\rm ss})  + w_2(-K_{\ell \ell'}^{\rm cs} - K_{\ell \ell'}^{\rm sc})
+w_3(-K_{\ell \ell'}^{\rm cs} - K_{\ell \ell'}^{\rm sc})
+w_4(K_{\ell \ell'}^{\rm cc} - K_{\ell \ell'}^{\rm ss})\bigg]\nonumber\\
&+\sin k\Delta \bigg[ w_1(-K_{\ell \ell'}^{\rm cc} + K_{\ell \ell'}^{\rm ss})  + w_2(K_{\ell \ell'}^{\rm cs} + K_{\ell \ell'}^{\rm sc})
+w_3(-K_{\ell \ell'}^{\rm cs} - K_{\ell \ell'}^{\rm sc})
+w_4(-K_{\ell \ell'}^{\rm cc} + K_{\ell \ell'}^{\rm ss})\bigg]\nonumber\\
&+\sin k\Sigma \bigg[ w_1(-K_{\ell \ell'}^{\rm cc} - K_{\ell \ell'}^{\rm ss})  + w_2(K_{\ell \ell'}^{\rm cs} - K_{\ell \ell'}^{\rm sc})
+w_3(K_{\ell \ell'}^{\rm cs} - K_{\ell \ell'}^{\rm sc})
+w_4(K_{\ell \ell'}^{\rm cc} + K_{\ell \ell'}^{\rm ss})\bigg]
\bigg\}.
\end{align}
Using the notation we have set up it would be straightforward, albeit algebraically involved, to obtain the analogous expression for a product of three or more spherical Bessel functions.


\bsp	
\label{lastpage}
\end{document}